\documentclass[
twocolumn,
prx,
amsmath,
amssymb,
longbibliography,
floatfix,
superscriptaddress,
aps]{revtex4-2}
\usepackage{amsmath}
\usepackage{amssymb}
\usepackage{xcolor}
\usepackage{physics}
\usepackage{graphicx}
\usepackage{tabularx}
\usepackage{booktabs}
\usepackage{dcolumn} 
\usepackage{bm}      
\usepackage{array}   
\usepackage{amsfonts}
\usepackage{url}
\usepackage{mathtools}
\usepackage[normalem]{ulem}
\usepackage{subfigure}

\usepackage{hyperref}
\hypersetup{
  colorlinks=true,
  linkcolor=cyan,
  citecolor=magenta,
  filecolor=magenta,
  urlcolor=cyan,
  runcolor=cyan}

\usepackage[ruled]{algorithm2e}
\usepackage[capitalize]{cleveref}

\def\>{\rangle}
\def\<{\langle}
\def\Tr{\mathrm{Tr}}
\def\Pr{\mathrm{Pr}}

\newcommand{\ketb}[2]{|{#1}\>\<#2|}

\newcommand{\kket}[1]{|#1\rangle\!\rangle}
\newcommand{\bbra}[1]{\langle\!\langle #1|}

\newcommand{\beq}{\begin{equation}}
\newcommand{\eeq}{\end{equation}}
\newcommand{\bal}{\begin{aligned}}
\newcommand{\eal}{\end{aligned}}
\newcommand{\bes} {\begin{subequations}}
\newcommand{\ees} {\end{subequations}}

\begin{document}

\title{QMCtwin: Master-Equation Simulation of Syndrome Statistics Beyond Pauli Noise}

\author{Tong Shen}
 \affiliation{Department of Electrical and Computer Engineering, University of Southern California, Los Angeles, California 90089, USA}
 \affiliation{Center for Quantum Information Science \& Technology, University of Southern California, Los Angeles, California 90089, USA}
 \affiliation{Quantum Elements, Inc., Westlake Village, 91361, CA, USA}
\author{Huo Chen}

 \affiliation{Quantum Elements, Inc., Westlake Village, 91361, CA, USA}
 \affiliation{Department of Physics, Harvard University, Cambridge, Massachusetts 02138, USA}
 
\author{Benchen Huang}
 \affiliation{AWS Worldwide Specialist Organization, Seattle,
98170, WA, USA}

\author{Tyler Takeshita}
 \affiliation{AWS Worldwide Specialist Organization, Seattle,
98170, WA, USA}

\author{Arian Vezvaee}
 \affiliation{Department of Electrical and Computer Engineering, University of Southern California, Los Angeles, California 90089, USA}
 \affiliation{Center for Quantum Information Science \& Technology, University of Southern California, Los Angeles, California 90089, USA}
 \affiliation{Quantum Elements, Inc., Westlake Village, 91361, CA, USA}

\author{Izhar Medalsy}
\affiliation{Quantum Elements, Inc., Westlake Village, 91361, CA, USA}

\author{Daniel A. Lidar}
 \affiliation{Department of Electrical and Computer Engineering, University of Southern California, Los Angeles, California 90089, USA}
 \affiliation{Center for Quantum Information Science \& Technology, University of Southern California, Los Angeles, California 90089, USA}
 \affiliation{Quantum Elements, Inc., Westlake Village, 91361, CA, USA}
 \affiliation{Department of Chemistry, University of Southern California, Los Angeles, California 90089, USA}
 \affiliation{Department of Physics and Astronomy, University of Southern California, Los Angeles, California 90089, USA}
 
\begin{abstract}
As quantum error correction moves toward large-scale experimental implementations, decoder performance increasingly depends on how faithfully hardware noise is translated into syndrome statistics. Standard stabilizer workflows achieve scalability by replacing device dynamics with stochastic Pauli or detector-error models, but this compression can discard coherent phase information, nonunital drift, continuous-time effects of always-on couplings, and correlations generated by simultaneous Hamiltonian and dissipative evolution. Here we present QMCtwin, a sign-problem-suppressed quantum Monte Carlo framework for master-equation simulation of QEC circuits, and apply it to a full syndrome-extraction round of a distance-$7$ rotated surface code with $97$ physical qubits. The open-system model includes realistic superconducting-device noise mechanisms such as relaxation, pure dephasing, coherent gate miscalibration, residual $ZZ$ crosstalk, and drive-qubit detuning. By directly estimating syndrome observables from the QMC-generated stochastic density matrix estimator, we compare the master-equation dynamics with their Pauli-twirled Clifford simulation counterparts. QMCtwin predicts syndrome-extraction biases and correlations between syndromes and proxies of logical-string-parity that are absent or strongly suppressed in the stochastic Pauli description. We introduce information-theoretic diagnostics that further quantify how information concerning syndromes versus string-parity proxies differs between the realistic master-equation simulation and the corresponding Pauli-twirled model. These results show that QMC-based master-equation digital twins can expose noise features hidden by conventional Pauli/Clifford noise models and provide a practical path toward more accurate decoder-facing syndrome models.

\end{abstract}

\maketitle

\section{Introduction}
Quantum error correction (QEC)~\cite{shor_scheme_1995,Steane:96a,lidar2013quantum} provides a systematic framework for suppressing noise and is widely regarded as the central requirement for achieving fault-tolerant quantum computation~\cite{Campbell:2017aa}. A central prediction of QEC theory is that, by encoding one logical qubit into many physical qubits, the logical error rate can be exponentially suppressed with increasing code size when the physical error rate is below a threshold~\cite{Aharonov:08,Gottesman:2013aa}. These predictions are increasingly supported by recent experiments demonstrating below-threshold scaling across multiple platforms~\cite{google2025quantum, bluvstein2026fault, reichardt2024demonstration,vezvaee2025surfacecodescalingheavyhex}, high-fidelity logical operations~\cite{paetznick2024demonstration, gupta2024encoding, chung2025fault,vezvaee2026demonstration}, and progress toward logical magic-state cultivation~\cite{rosenfeld2025magic} and distillation~\cite{sales2025experimental}.

As hardware improves, the emphasis is shifting from suppressing physical error rates alone to decoding the syndrome information produced by the hardware~\cite{he2025experimental}. Each QEC round measures stabilizers, producing a noisy and incomplete syndrome record from which a classical decoder must infer the most likely error history and choose a recovery operation or Pauli-frame update~\cite{dennis2002topological, fowler2012surface}. While small codes can be decoded by table lookup, decoding becomes a difficult inference problem at scale. This is especially true for large topological codes and higher-rate quantum LDPC codes~\cite{breuckmann2021quantum}, where the decoder must infer likely error histories from a finite set of parity-check outcomes, and where realistic hardware noise may be coherent, correlated, and time-distributed across the syndrome-extraction circuit. Thus, logical performance depends not only on the physical error rates of the device, but also on the accuracy of the decoder's statistical model of how physical noise maps into syndrome patterns. If this model differs substantially from the true device noise, the decoder can assign incorrect likelihoods to candidate error histories and make systematically suboptimal recovery decisions~\cite{tuckett2020fault,higgott2023improved}.

A common route to scalable decoding is to approximate device noise by an effective stochastic Pauli or detector-error model (DEM)~\cite{gidney2021stim,Higgott2023sparse,EisertarXiv2024}. Coherent gate miscalibrations, relaxation, and dephasing are often represented by Pauli error channels inserted at circuit locations, for example after Pauli twirling~\cite{dur2005standard,geller2013efficient}; reset and measurement faults are then represented by corresponding preparation, readout, or detector-error channels. This enables syndrome circuits to be simulated efficiently with stabilizer/Clifford simulators~\cite{gottesman1997stabilizer, aaronson2004improved} and compiled into DEMs for fast decoding. However, this compression can remove information that matters for decoding under realistic noise~\cite{schwartzmannowik2025modeling,katsuda2024simulation,ni2024superconducting,miller2025efficient,myers2025general,tuloup2026pauliframe,leblond2025logical,harper2026crosstalk,barone2025colorcode,hines2026simulating,takou2025coherent}: 
\begin{enumerate}
  \item At a fixed circuit location, Pauli twirling diagonalizes the Pauli-transfer representation of the local channel, erasing coherent phase information and phase-sensitive interference; for nonunital channels such as amplitude damping, it also removes affine drift terms.
  \item Although any finite-time noise process can be Pauli-twirled as a channel, the syndrome distribution of a circuit with continuously acting coherent terms, such as residual $ZZ$ crosstalk or detuning during driven pulses, is generally not reproduced by inserting a sequence of local, independent, per-gate Pauli errors.
  \item In the device model considered here, coherent Hamiltonian terms and dissipative processes act concurrently. For example, a miscalibrated gate can interact with crosstalk, detuning, relaxation, dephasing, or spectator-qubit dynamics, producing syndrome patterns whose likelihoods need not match those predicted by combining individually twirled error channels.
\end{enumerate}
These limitations motivate syndrome simulations under realistic noise that preserve coherent, dissipative, and correlated dynamics before compressing them into decoder-facing statistics.

A natural way to preserve these effects is to simulate the syndrome-extraction circuit using an open-system master equation that directly captures coherent control errors, dissipation, residual couplings, and their interplay during the circuit. The challenge is computational cost. Dense master-equation simulation requires $O(4^n)$ memory for $n$ qubits, and even sparse local Liouvillian time evolution acts on an exponentially large state space, becoming prohibitive at modest system sizes. Quantum-trajectory methods reduce memory by sampling pure-state trajectories~\cite{dalibard1992wave,Johansson:2012aa,chen2020hoqst}, but each trajectory remains a full $2^n$-dimensional state vector.  Moreover, standard Markovian stochastic unravelings rely on non-negative instantaneous jump rates; time-local master equations with temporarily negative rates~\cite{Rivas:2014aa} require modified non-Markovian trajectory schemes and can introduce additional sampling difficulties~\cite{becker2023quantum}. Tensor‑network methods~\cite{vidal2003efficient, verstraete2004matrix, zwolak2004mixed} can be powerful when the dynamics remain compressible; for 2D QEC circuits with realistic couplings and open‑system noise, that compressibility can degrade, requiring approximations that may obscure correlated and phase‑sensitive noise signatures~\cite{weimer2021simulation}.

Against this backdrop, we build on our recent real-time, sign-problem-suppressed quantum Monte Carlo (QMC) framework~\cite{shen2025real}, which stochastically compresses and evolves open-system density matrices and makes master-equation-faithful simulation feasible at much larger scales. In this work, we apply that framework to an experiment-scale syndrome-extraction circuit setting, performing, to our knowledge, one of the largest open-system master-equation simulations: a full round of a distance-$7$ rotated surface code with $97$ physical qubits, comparable in scale to recent experimental surface‑code memory demonstrations~\cite{google2025quantum}. We call our QMC framework adapted to experiment-scale syndrome-extraction circuits ``QMCtwin''. We apply QMCtwin to a hardware-motivated superconducting noise model that includes relaxation, dephasing, coherent miscalibration, residual crosstalk, and detuning. The resulting stochastic density-matrix simulation reveals syndrome-extraction biases and syndrome-to-logical-string-parity correlations that are absent or strongly suppressed in the reference Pauli-twirled Clifford simulation. These results establish realistic open-system noise simulation as a practical route to identify decoder-relevant structure before it is compressed into decoder-facing statistical models.

This paper is organized as follows. In \cref{sec:prior} we summarize relevant prior work and situate our own work in this context. In \cref{sec:methods}, we introduce the toggling-frame master-equation formulation and QMC estimator. In \cref{sec:model}, we describe the surface-code circuit and superconducting-device noise model. In \cref{sec:results}, we present QMC diagnostics, syndrome-extraction biases, and information-theoretic syndrome-to-string-parity comparisons. We conclude with implications for decoder-facing noise models. The appendix contains additional technical details in support of the main text.

\section{Prior work}
\label{sec:prior}

There has been significant recent work on QEC beyond Pauli-stochastic models. The common message is that Pauli twirling can misestimate syndrome statistics, thresholds, and logical failure rates.

Most closely related to the present work are explicit open-system studies, but so far these have remained significantly smaller than our $97$ qubit simulation. Schwartzman-Nowik \emph{et al.}~\cite{schwartzmannowik2025modeling} analyzed the five-qubit code under Lindblad dynamics with coherent, dissipative, and two-qubit crosstalk terms, clarifying where composite-channel and Pauli approximations fail. Katsuda \emph{et al.}~\cite{katsuda2024simulation} performed a full realistic-noise simulation of a distance-$5$ rotated surface code with $49$ qubits, reducing the simulated register to $26$ qubits by delaying syndrome measurements and then fitting an effective stochastic model. Ni \emph{et al.}~\cite{ni2024superconducting} introduced a Hamiltonian-to-QEC workflow for superconducting processors that propagates correlated unitary errors from a device Hamiltonian to logical-memory performance and gradients for design optimization. 

A second body of work simulates beyond-Pauli QEC circuits more indirectly and approximately. Miller \emph{et al.}~\cite{miller2025efficient} developed an approximate simulation method for Clifford circuits with arbitrary small sparse Lindbladian errors, and applied it to rotated-surface-code syndrome extraction at distances $3,5,7,9,11$ as well as random 225-qubit circuits. Myers \emph{et al.}~\cite{myers2025general} used stratified importance sampling to simulate general noise within the stabilizer formalism, making many nonunitary channels nearly as cheap as Pauli noise while extending direct rotated-surface-code studies beyond the Pauli approximation. Tuloup and Ayral~\cite{tuloup2026pauliframe} introduced the Pauli Frame Sparse Representation with truncation and shot-efficient importance sampling, finding that Pauli twirling can overestimate coherent-noise thresholds by about a factor of four in circuit-level threshold studies up to distance $9$. LeBlond \emph{et al.}~\cite{leblond2025logical} used a quasi-probability decomposition and phase-sensitive Clifford simulation to study a trapped-ion-inspired surface-code memory under mixed coherent and stochastic circuit-level noise, extracting logical channels up to distance $11$. 

Tensor-network approaches have also moved beyond Pauli approximations: Harper \emph{et al.}~\cite{harper2026crosstalk} used hybrid stabilizer-tensor-network methods to simulate coherent $ZZ$ crosstalk during surface-code syndrome extraction for distances $3,5,7,9$, while Barone \emph{et al.}~\cite{barone2025colorcode} used tree-tensor-network methods together with quantum trajectories to estimate color-code thresholds under coherent over-rotations and amplitude damping up to $73$ qubits. 

Hines \emph{et al.}~\cite{hines2026simulating} introduced a perturbative generator-to-DEM pipeline for small Markovian circuit-level errors represented by sparse elementary error generators, showing that DEMs built from the non-Pauli model predict detection-history distributions for two rounds of distance-$3$ surface-code syndrome extraction much more accurately than Pauli-twirled surrogates, and enabling threshold and logical-rate studies for surface-code memory, bivariate bicycle codes, and magic-state cultivation. Takou and Brown~\cite{takou2025coherent} then showed that coherent-error interference and hyperedges absent from Pauli-twirled models can be inferred from syndrome histories themselves, so that experimentally estimated DEMs can be used in both stochastic and coherent regimes. Closely related syndrome-data approaches estimate DEM events, decoding graphs, hypergraphs, or correlated matching weights directly from measured correlations rather than from a microscopic hardware model~\cite{blumekohout2025estimating,remm2025experimentally,takou2025graphs}. 

These methods generally begin from circuit-level error channels or sparse decompositions rather than directly integrating the time-dependent Lindblad dynamics. By contrast, QMCtwin directly simulates a full distance-$7$ rotated-surface-code syndrome-extraction round under a superconducting-device Lindblad model containing nonunital relaxation, pure dephasing, coherent pulse-amplitude miscalibration, drive-qubit detuning, and spatially inhomogeneous residual $ZZ$ crosstalk, and it focuses on the intermediate decoder-facing layer of syndrome statistics before full DEM construction or logical-error-rate decoding. In this sense, the present work is complementary to DEM-oriented and logical-channel-oriented approaches: rather than immediately compressing the dynamics to detector events or logical failure counts, we quantify what diagnostic structure survives or is lost when a realistic open-system simulation is replaced by a Pauli-twirled Clifford baseline.

\section{Methods}
\label{sec:methods}

\subsection{Frame transformations}

In the laboratory frame, the Schr\"odinger-picture Markovian master equation of a system with density matrix $\rho_{\rm lab}(t)$ can be written in Lindblad form~\cite{gorini1976completely,lindblad1976generators,Breuer:book} as
\beq
    \dot\rho_{\rm lab}(t) = -i[H_{\rm lab}(t), \rho_{\rm lab}(t)] + \frac{1}{2}\sum_\ell \gamma_\ell(t) \mathcal{D}[L_\ell^{\rm lab}] \rho_{\rm lab}(t),
    \label{eq:lindblad_lab}
\eeq
where
\beq
    \mathcal{D}[L]\rho := 2L\rho L^{\dagger} - \{L^{\dagger}L, \rho \} .
\eeq
Here $H_{\rm lab}(t)$ contains the physical system Hamiltonian and applied control fields in the laboratory frame, while the Lindblad operators $L_\ell^{\rm lab}$ describe processes due to the coupling with the environment, with non-negative rates $\gamma_\ell(t)$.  

For the pulse-level device model used below, we transform this equation to a rotating control frame. Let $R(t)$ denote the chosen rotating-frame transformation (see Appendix~\ref{app:RF} for details), and define
\beq
    \rho(t)=R^{\dagger}(t)\rho_{\rm lab}(t)R(t).
\eeq
The Hamiltonian and Lindblad operators in this frame are
\beq\bal
    H(t)
    &=
    R^{\dagger}(t)H_{\rm lab}(t)R(t)
    -
    iR^{\dagger}(t)\dot R(t),\\
    L_\ell(t)
    &=
    R^{\dagger}(t)L_{\ell}^{\rm lab}R(t).
\eal
\eeq
In this rotating control frame, the master equation has the same form,
\beq
    \dot\rho(t)
    =
    -i[H(t),\rho(t)]
    +
    \frac{1}{2}\sum_\ell \gamma_\ell(t)
    \mathcal{D}[L_\ell(t)]\rho(t).
    \label{eq:lindblad_control}
\eeq
In the simulations we describe below, the Lindblad operators are specified in this rotating frame. They may acquire frame-dependent phases; when this time-dependence is only a scalar phase, it is immaterial to the dissipator, and we write the operators simply as $L_\ell$. Any time dependence that changes the dissipator is kept explicitly as $L_\ell(t)$.

Time-local descriptions of non-Markovian dynamics can be written in the same form with temporarily negative rates, but the generator is then no longer in Lindblad form with nonnegative rates, and the evolution is not CP-divisible~\cite{Rivas:2014aa}. An arbitrary such time-local generator need not produce a completely positive dynamical map~\cite{breuer2016colloquium}. The simulations reported below use non-negative Markovian rates.

For the syndrome-extraction circuit considered here, the ideal circuit evolution is known and can be tracked exactly. What matters is the evolution relative to this known ideal circuit, i.e., how coherent and incoherent noise propagates into syndrome statistics. We therefore remove the perfect gate dynamics by moving to a toggling frame defined by the ideal control evolution. In this frame, error-free evolution corresponds to the identity process, and the master equation evolves only the residual error dynamics as transformed by the ideal circuit. 

In the rotating control frame, we first split the Hamiltonian as
\beq
    H(t)=H_{\rm ideal}(t)+H_{\rm rest}(t),
    \label{eq:Hsplit}
\eeq
where $H_{\rm ideal}(t)$ generates the intended gates, and $H_{\rm rest}(t)$ contains coherent imperfections, residual couplings, detunings, and other Hamiltonian terms not included in the ideal circuit. Let $U_{\rm ideal}(t)$ be the solution of
$\dot U_{\rm ideal}(t)=-iH_{\rm ideal}(t)U_{\rm ideal}(t)$, $U_{\rm ideal}(0)=I$. The toggling-frame density matrix is defined by
\beq
    \tilde{\rho}(t)=U_{\rm ideal}^{\dagger}(t) \rho(t) U_{\rm ideal}(t),
    \label{eq:rho_tilde_def}
\eeq
and it obeys
\beq
    \dot{\tilde{\rho}}
    = -i[\tilde{H}(t),\tilde{\rho}]
    + \frac{1}{2}\sum_\ell \gamma_\ell(t)     \mathcal{D}[\tilde{L}_{\ell}(t)]\tilde{\rho}(t),
    \label{eq:lindblad_toggling_general}
\eeq
with
\beq
\begin{aligned}
    \tilde{L}_{\ell}(t) &= U_{\rm ideal}^{\dagger}(t)L_{\ell}(t)U_{\rm ideal}(t),
    \\
    \tilde{H}(t) &= U_{\rm ideal}^{\dagger}(t) H_{\rm rest}(t) U_{\rm ideal}(t).
    \label{eq:control_cancelled}
\end{aligned}
\eeq
Thus, the toggling frame removes the explicit action of the ideal gates. The remaining coherent and dissipative error mechanisms are instead rotated by the ideal circuit evolution. This frame therefore describes the error-only dynamics as seen by the QEC circuit. From this point onward, unless explicitly labeled ``lab'', all un-tilded quantities are expressed in the rotating control frame, while tilded quantities are expressed in the ideal-circuit toggling frame.

We solve \cref{eq:lindblad_toggling_general} using the real-time QMC algorithm introduced in Ref.~\cite{shen2025real}. This algorithm is summarized in Appendix~\ref{appendix-QMC}. 

To evaluate observables in the rotating control frame, the observable can be rotated into the ideal-circuit toggling frame:
\begin{align}
    \expval{O(t)} &= \Tr[O(t) \rho(t)] \nonumber\\
    & = \Tr[U^{\dagger}_{\rm ideal}(t) O(t) U_{\rm ideal}(t) \tilde{\rho}(t)] = \Tr[\tilde{O}(t) \tilde{\rho}(t)],
    \label{eq:toggling_frame_expect}
\end{align}
where $\tilde{O}(t) = U_{\rm ideal}^{\dagger}(t) O(t) U_{\rm ideal}(t)$ is the toggling-frame observable. A laboratory-frame observable $O_{\rm lab}(t)$ should first be expressed in the rotating control frame as $O(t)=R^{\dagger}(t)O_{\rm lab}(t)R(t)$.

\subsection{Syndrome readout}

\begin{figure}
    \centering
    \includegraphics[width=1.0\linewidth]{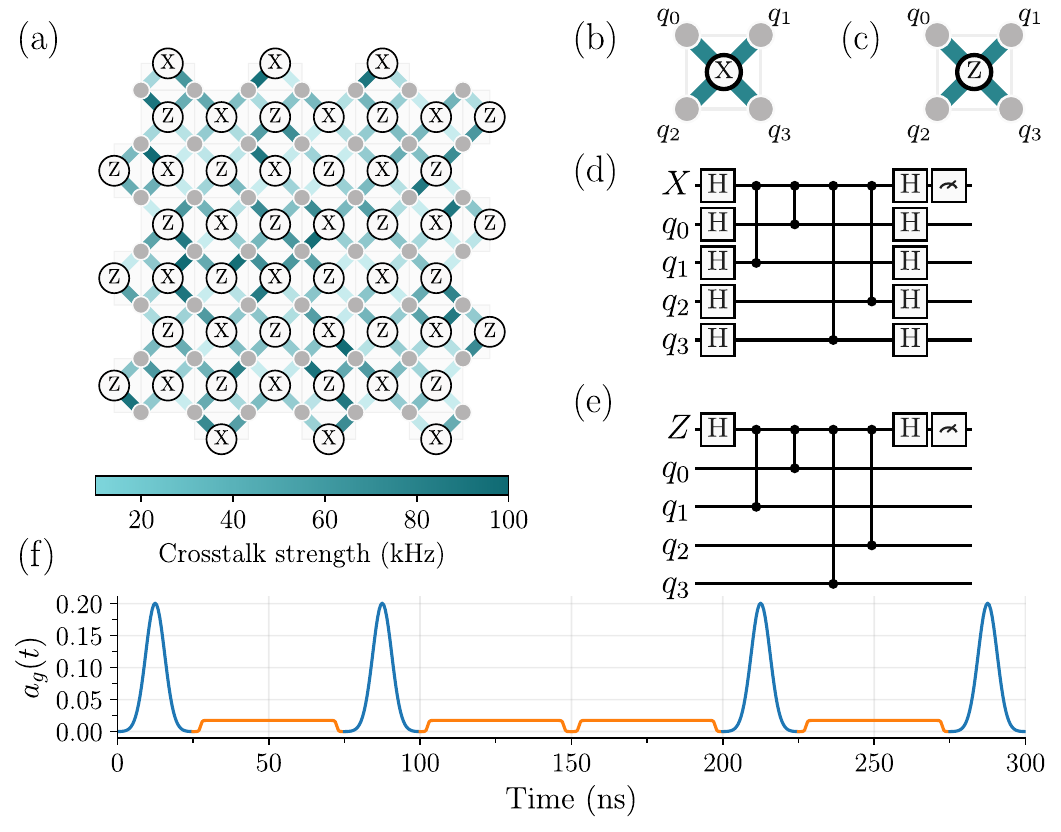}
    \caption{Surface-code layout and syndrome-extraction circuits used in the distance-$7$ simulation. (a) Rotated surface-code layout with $49$ data qubits (grey disks) and $48$ measurement qubits (circles with X or Z). Measurement qubits are labeled by the stabilizer type they measure, $X$ or $Z$, and edges indicate nearest-neighbor couplings $J_{ij}/2\pi$; the color scale shows the sampled residual $ZZ$ crosstalk strength.  (b-c) Representative four-body $X$- and $Z$-type stabilizer neighborhoods, respectively. (d-e) Gate schedules for measuring $X$- and $Z$-type stabilizers during the syndrome-extraction round. A box with an H denotes a Hadamard gate, and a solid vertical line with filled endpoints denotes a CZ-equivalent entangling gate. (f) Representative layer-level pulse envelope for one syndrome-extraction round. Blue pulses denote 25 ns Gaussian single-qubit gates, and orange pulses denote 50 ns unipolar-sigmoid two-qubit gates.}
    \label{fig:surface_code}
\end{figure}

Let $\mathcal D$ denote the set of data qubits, let $\mathcal A$ denote the set of measurement ancilla qubits, and let $\mathcal C_X$ and $\mathcal C_Z$ denote the sets of $X$-type and $Z$-type stabilizer check labels measured in the syndrome-extraction round [see \cref{fig:surface_code}(a)]. 
We write $\mathcal C = \mathcal C_X\cup \mathcal C_Z$ for the full set of check labels. For each check $k\in \mathcal C$, let $a(k)\in\mathcal A$ be the measurement ancilla qubit associated with that check. The intended data stabilizer for check $k$ is denoted by $S_k$, with
\beq
     S_k=
    \begin{cases}
    S_k^X=\prod_{i\in\partial k} X_i, & k\in\mathcal C_X,\\
    S_k^Z=\prod_{i\in\partial k} Z_i, & k\in\mathcal C_Z,
   \end{cases},
    \label{eq:parity_check}
\eeq
where $\partial k$ denotes the set of data qubits neighboring check $k$. 

At the end of the round, at time $t_f$, the hardware readout is a computational-basis measurement of each measurement qubit. Since the rotating-frame transformation $R(t)$ is a product of local $Z$ rotations (see Appendix~\ref{app:RF}), the computational-basis Pauli $Z_{a(k)}$ has the same operator form in the laboratory and rotating control frames.

For a full syndrome outcome $\mathbf m=\{m_k\}_{k\in\mathcal C}$ with $m_k\in\{-1,+1\}$, 
the final readout projector in the rotating control frame is
\beq
    M_{\mathbf m}^{\rm rd}
    =
    \prod_{k\in \mathcal{C}}
    \frac{I+m_k Z_{a(k)}}{2}.
\label{eq:readout_projector}
\eeq
The corresponding toggling-frame projector is
\beq
    \tilde M_{\mathbf m}^{\rm rd}(t)
    =
    U_{\rm ideal}^{\dagger}(t) M_{\mathbf m}^{\rm rd} U_{\rm ideal}(t),
\eeq
and the syndrome probability is
\beq
    \Pr(\mathbf m) = \Tr[\tilde M_{\mathbf m}^{\rm rd}(t_f) \tilde{\rho}(t_f)].
    \label{eq:prob_from_dm}
\eeq

The connection between the readout operators and the intended stabilizers is fixed by the ideal syndrome-extraction circuit. Let $\rho_{\mathcal A}^{0}$ denote the prepared initial state of the measurement qubits at the start of the round. In the simulations below, $\rho_{\mathcal A}^{0}=\bigotimes_{a\in\mathcal A}\ket{0}_a\!\bra{0}$, with any subsequent ancilla basis-preparation gates included in $U_{\rm ideal}(t_f)$. We choose the syndrome-label convention so that a $+1$ readout of the measurement ancilla corresponds to a $+1$ eigenvalue of the intended stabilizer $S_k$.
With these conventions, the noiseless ideal circuit implements a simultaneous projective measurement of the commuting stabilizers. The induced data observable associated with readout qubit $a(k)$ satisfies
\beq
    \Tr_{\mathcal A} \left[
        \left(I_{\mathcal D}\otimes\rho_{\mathcal A}^{0}\right)
        U_{\rm ideal}^{\dagger}(t_f)
        Z_{a(k)}
        U_{\rm ideal}(t_f)
    \right]
    =
    S_k,
    \label{eq:readout_stabilizer_relation}
\eeq
and the corresponding joint data positive operator-valued measure (POVM) satisfies
\beq
    \Tr_{\mathcal A} \left[
        \left(I_{\mathcal D}\otimes\rho_{\mathcal A}^{0}\right)
        \tilde M_{\mathbf m}^{\rm rd}(t_f)
    \right]
    =
    \prod_{k\in\mathcal C}
    \frac{I+m_k S_k}{2}.
    \label{eq:readout_projector_stabilizer_relation}
\eeq

If the measurement process itself is noisy, $M_{\mathbf m}^{\rm rd}$ in \cref{eq:readout_projector} should be replaced by the appropriate rotating-control-frame POVM element for outcome $\mathbf m$ before applying the toggling-frame rotation.
In the simulations below, we do not include separate state-preparation, reset, or measurement noise; reported syndrome probabilities correspond to ideal projective readout of the final noisy state. This choice isolates the propagation of coherent and dissipative circuit-level noise into the pre-readout syndrome distribution. In comparisons with hardware, final readout-assignment errors could be included explicitly through a calibrated POVM or response matrix, and in some settings partially mitigated using measurement-error-mitigation methods~\cite{Nation:2021aa}.

\section{Model and Simulation Details}
\label{sec:model}

As an experiment-scale proof of concept, we simulate a single syndrome-extraction round of a distance-$7$ rotated surface code with $97$ physical qubits using our QMCtwin master-equation solver. The simulated device is motivated by superconducting transmon qubits~\cite{transmon-invention,Schreier2008,Paik2011,Blais2021,tripathi2024modeling}. The Hamiltonian entering \cref{eq:lindblad_control} is written as
\beq
    H(t)=H_{\rm static}+H_{\rm gate}(t),
\eeq
where the static residual contribution is
\beq
    H_{\rm static} = -\frac{1}{2}\sum_{i=1}^{n} \delta\omega_{i} \sigma_i^{z} + \sum_{\<i,j\>} J_{ij} \sigma_i^{z}\sigma_j^{z}.
\eeq
Here $\delta\omega_{i}$ is the residual static detuning of qubit $i$ in the rotating control frame, and $J_{ij}$ is the residual always-on $ZZ$ coupling coefficient between neighboring qubits $i$ and $j$. In the simulations reported here, the drive-qubit detuning $\Delta_g$ is included in the driven-pulse Hamiltonian, and we set $\delta\omega_i=0$ unless explicitly stated otherwise.

Let $\mathcal G(t)$ denote the intended gate set and let $H_g^{\rm ideal}(t)$ denote the corresponding ideal pulse Hamiltonian for gate $g$. The ideal control Hamiltonian is
\beq
    H_{\rm ideal}(t)
    =
    \sum_{g\in\mathcal G(t)}H_g^{\rm ideal}(t).
\eeq
For the toggling-frame evolution, we write the corresponding implemented control Hamiltonian as
\beq
    H_{\rm gate}(t)=\sum_{g\in\mathcal G(t)} H_g(t) .
\eeq
The residual Hamiltonian entering \cref{eq:control_cancelled} is
\beq
    H_{\rm rest}(t)
    =
    H_{\rm static}
    +
    \sum_{g\in\mathcal G(t)}
    \left[
        H_g(t)-H_g^{\rm ideal}(t)
    \right].
\eeq
Drive detuning, coherent under-rotation, and pulse-shape imperfections are included in $H_g(t)-H_g^{\rm ideal}(t)$.

Dissipation is modeled by local relaxation and pure-dephasing channels. With the dissipators in \cref{eq:lindblad_control}, we use
\beq
    L_{1,i}=\sigma_i^{-},\quad
    \gamma_{1,i}=\frac{1}{T_{1,i}}\ge 0,
\eeq
for energy relaxation, and
\beq
    L_{\phi,i}=\sigma_i^{z}, \quad
    \gamma_{\phi,i}=\frac{1}{2T_{\phi,i}}\ge 0,
\eeq
for pure dephasing. 
The effective transverse relaxation time is 
\beq
\label{eq:T2}
    \frac{1}{T_{2,i}} = \frac{1}{2T_{1,i}} + \frac{1}{T_{\phi,i}}.
\eeq

To anchor the model in experimentally realistic parameter ranges, we use characterization data from \texttt{ibm\_miami}. 
For the simulated device, 
$T_1$, $T_\phi$, and residual $ZZ$ couplings are sampled from these experimentally motivated ranges. The coherent pulse-amplitude miscalibration is fixed to $\delta_g=0.1\%$. In the detuning results described below, we take the drive-qubit detuning $\Delta_g/2\pi$ to be the same for all single-qubit pulses in a given simulation, and sweep it over the stated values.
The resulting noise model is summarized in \cref{tab:noise_model}, together with the Pauli-twirled reference error parameters used for the Clifford simulations in \cref{sec:results} below. Additional characterization data motivating these parameter ranges are shown in Appendix~\ref{app:additional-modeling}.

\begin{table}[t]
\centering
\setlength{\tabcolsep}{4.5pt}
\renewcommand{\arraystretch}{1.12}
\begin{tabular}{@{}ccc@{}}
\toprule
Noise source & ME parameter range & PT error probability \\
\midrule
$T_1$ &
$150\text{--}300\mu{\rm s}$ &
$2\times10^{-5}\text{--}8\times10^{-5}$ \\


$T_\phi$ &
$90\text{--}165\mu{\rm s}$ &
$10^{-4}\text{--}10^{-3}$ \\

$\Delta_g/2\pi$ &
$-50\text{--}50{\rm kHz}$ &
--- \\

$J_{ij}/2\pi$ &
$10\text{--}100{\rm kHz}$ &
$10^{-8}\text{--}10^{-3}$ \\

$\delta_g$ &
$0.1\%$ &
$\sim 10^{-6}$ \\
\bottomrule
\end{tabular}
\caption{Noise parameters used in the master-equation (ME) simulations and the Pauli-twirled (PT) Clifford error probabilities. PT entries are representative order-of-magnitude Pauli-error probability ranges for the gate and idle durations used in the Stim circuit; the probabilities actually inserted into Stim are defined in Appendix~\ref{app:Pauli-twirling}.}
\label{tab:noise_model}
\end{table}

Time-dependent gates are modeled as driven control terms with virtual-$Z$ frame updates. For driven single-qubit pulses, we write
\beq
        H_{\rm gate}^{1q}(t) =
    \sum_{g\in \mathcal{G}_{1q}(t)}
    \left(
        {\rm Re}[\Omega_g(\tau)] H_{g,x}
        +
        {\rm Im}[\Omega_g(\tau)] H_{g,y}
    \right),
    \eeq
where
\beq
    \label{eq:Omega}
    \Omega_g(\tau)
    =
    a_g(\tau)\exp\left[i(\Delta_g \tau+\phi_g)\right],    
\eeq
and where $\mathcal{G}_{1q}(t)$ is the set of active driven single-qubit pulses, $\tau$ is the local time within the pulse, $H_{g,x}$ and $H_{g,y}$ are the normalized quadrature generators, chosen so that $\int a_g^{\rm ideal}(\tau)d\tau$ is the intended rotation angle. For a pulse acting on qubit $i_g$, this corresponds to $H_{g,x}=X_{i_g}/2$ and $H_{g,y}=Y_{i_g}/2$ in our convention. The pulse envelope $a_g(\tau)$ is real-valued, and $\phi_g$ is the drive phase, including the phase accumulated from prior virtual-$Z$ frames on the driven qubit. 
Longitudinal two-qubit $R_{ZZ}$ pulses are modeled separately as
\beq
    H_{\rm gate}^{2q}(t)
    =
    \sum_{g\in\mathcal{G}_{ZZ}(t)}
    \Omega_{ZZ,g}(\tau) H_{ZZ,g},
\eeq
where $\mathcal G_{ZZ}(t)$ is the set of active longitudinal two-qubit pulses at time $t$, and with $H_{ZZ,g}$ chosen so that, together with the associated virtual-$Z$ frame updates and an irrelevant global phase, the ideal pulse implements the desired CZ-equivalent entangling operation. Thus, for a target $R_{ZZ}$ rotation on qubits $i_g$ and $j_g$, the generator $H_{ZZ,g}$ is proportional to $Z_{i_g}Z_{j_g}$; e.g., for
\beq
\label{eq:RZZ}
R_{ZZ}(\theta)=\exp(-i\theta Z_iZ_j/2),
\eeq
we take
\beq
    H_{ZZ,g}=\frac{1}{2}Z_{i_g}Z_{j_g},\quad \int \Omega_{ZZ,g}^{\rm ideal}(\tau)d\tau=\theta .
\eeq
The full gate Hamiltonian is
\beq
    H_{\rm gate}(t)=H_{\rm gate}^{1q}(t)+H_{\rm gate}^{2q}(t).
\eeq
For a driven single-qubit pulse $g$ acting on qubit $i$,  the detuning is defined as
\beq
    \Delta_g = \omega_{c,g}-\omega_{q,i},
\eeq
where $\omega_{c,g}$ is the carrier angular frequency of the drive and $\omega_{q,i}$ is the qubit transition angular frequency. This sign convention is the one used in the pulse phase in \cref{eq:Omega}. We model coherent gate miscalibration by scaling each nonvirtual physical pulse envelope by the same relative factor. For driven single-qubit pulses,
\beq
    a_g(\tau)=(1-\delta_g)a_g^{\rm ideal}(\tau),
\eeq
and for longitudinal two-qubit pulses,
\beq
    \Omega_{ZZ,g}(\tau)=(1-\delta_g)\Omega_{ZZ,g}^{\rm ideal}(\tau),
\eeq
with miscalibration $\delta_g=0.1\%$. Virtual-$Z$ frame updates are not assigned an independent amplitude error in this model.

The surface code consists of a two-dimensional lattice of data qubits and measurement qubits~\cite{fowler2012surface}. 
For the distance-$7$ rotated surface code used here, $d^2=49$ data qubits are coupled to $d^2-1=48$ measurement qubits, giving $97$ physical qubits in total [\cref{fig:surface_code}(a)]. 
Each measurement qubit is assigned to either an $X$- or $Z$-type stabilizer as defined in \cref{eq:parity_check}.
Bulk checks have weight four, as illustrated in \cref{fig:surface_code}(b-c), while boundary checks have the corresponding reduced weight.

We use a CZ-equivalent stabilizer-measurement circuit. For an $X$-type check, Hadamard basis changes are applied to the measurement qubit and to the neighboring data qubits before and after the entangling operations, mapping the $X$-parity measurement to a $Z$-basis parity measurement [\cref{fig:surface_code}(d)]. 
For a $Z$-type check, the neighboring data qubits remain in the computational basis, while the measurement qubit is prepared and read out through Hadamard rotations [\cref{fig:surface_code}(e)]. 

\begin{figure}[t]
    \centering
    \includegraphics[width=0.8\linewidth]{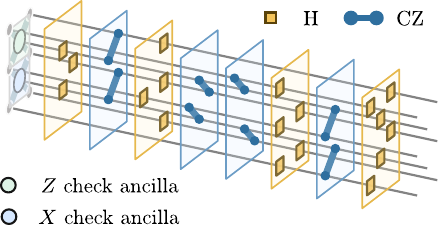}
    \caption{Representative two-plaquette schedule used to construct the distance-$7$ syndrome-extraction round in the simulation. The schedule is obtained by applying the $X$ and $Z$ checks in \cref{fig:surface_code}(d,e) across the layout in \cref{fig:surface_code}(a), with the single-qubit basis-change gates and the two-qubit entangling gates grouped into conflict-free layers.}
    \label{fig:full_schedule}
\end{figure}

The full distance-$7$ syndrome-extraction round is obtained by applying these local $X$- and $Z$-check circuits across the layout of \cref{fig:surface_code}(a). 
The gates are grouped into eight conflict-free layers so that no two simultaneously active operations share a qubit: four single-qubit layers containing the Hadamard-basis changes required by the local checks, and four two-qubit layers containing the CZ-equivalent $R_{ZZ}(\pi/2)$ gates associated with the check-data interactions. 
A representative two-plaquette portion of this schedule is shown in \cref{fig:full_schedule}. 
The complete distance-$7$ round contains $228$ single-qubit gates and $168$ two-qubit entangling gates and, with $25~{\rm ns}$ single-qubit layers and $50~{\rm ns}$ two-qubit layers, has total duration $300~{\rm ns}$.

At the pulse level, each Hadamard gate is compiled into one physical $R_X(\theta_H)$ rotation driven by a $25$ ns Gaussian pulse, with the required $Z$ rotations absorbed into virtual-$Z$ frame updates~\cite{McKay:2017tv,vezvaee2025virtual}. The angle $\theta_H$ is the physical rotation angle used in this compilation, which we set equal to $\pi/2$. Each two-qubit entangling gate is implemented by a CZ-equivalent $R_{ZZ}(\pi/2)$ rotation [\cref{eq:RZZ}], driven by a $50$ ns unipolar-sigmoid pulse, with associated local $Z$ phases also treated virtually. Since all gates of the same type within a layer share the same pulse shape and duration, \cref{fig:surface_code}(f) shows a representative layer-level pulse envelope. For each layer, the corresponding envelope is applied only to the active qubits or qubit pairs specified by the stabilizer-measurement circuits in \cref{fig:full_schedule}.

We hold the sampled $T_1$, $T_\phi$, and residual-$ZZ$ parameters, as well as the fixed pulse-amplitude miscalibration $\delta_g$, constant across the detuning scan, and sweep the uniform drive-qubit detuning over $\Delta_g/2\pi=\{-50,-25,0,+25,+50\}$ kHz. Each QMC run was initialized with the $\ket{0\dots 0}$ state and initial diagonal walker count $N_{\rm diag}=10^7$. For each detuning, results are averaged over five independent QMC runs. Error bars are obtained by bootstrap resampling over these independent runs and are semi-quantitative uncertainty estimates for this small ensemble. The simulations were performed on Amazon EC2 Hpc7a instances managed with AWS ParallelCluster, requiring on average $75$ minutes per run on a single $96$-vCPU instance. 

The Clifford simulations were conducted with Stim~\cite{gidney2021stim} using the stochastic noise model in \cref{tab:noise_model}, with $N_{\rm Stim}=10^7$ shots for each setting. 
The Pauli-twirled reference model contains no separate stochastic channel for drive-qubit detuning: detuning is retained only in the master-equation pulse Hamiltonian because it acts during driven rotations and is correlated with the intended coherent control. Representing it as an independent stochastic Pauli fault at a circuit location would not reproduce the same pulse-level dynamics. 
Thus, direct same-parameter ME-Clifford comparisons are made only at $\Delta_g=0$; at nonzero detuning,  where applicable, we use the same Stim model only as a fixed Pauli-twirled predictor evaluated on ME-estimated reference distributions.

\section{Results}
\label{sec:results}

\begin{figure}
    \centering
    \includegraphics[width=1.0\linewidth]{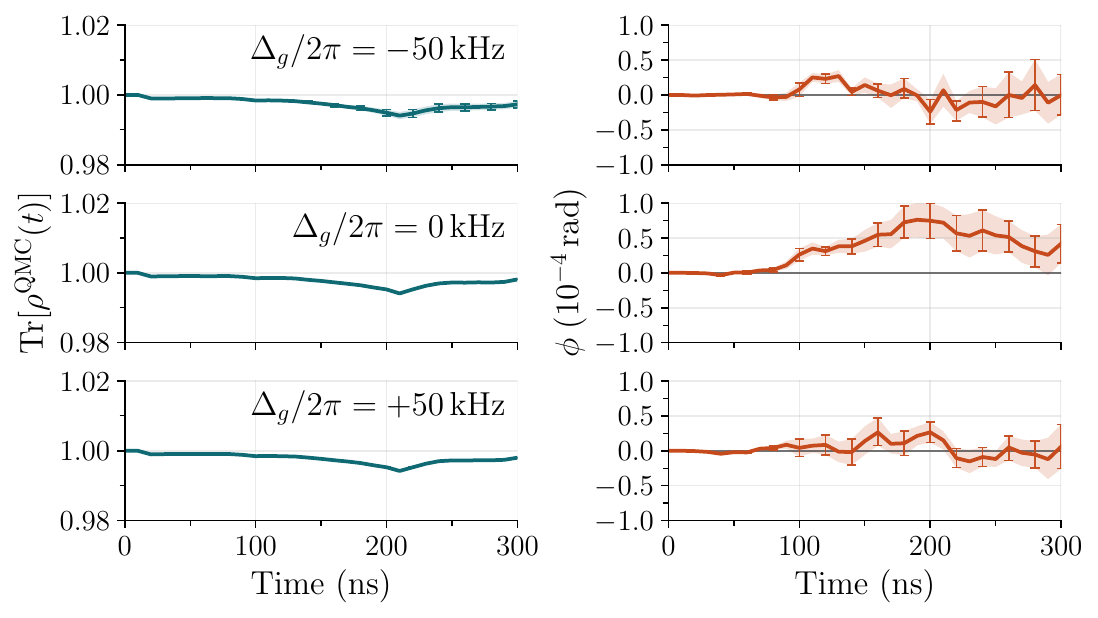}
    \caption{QMC trace and phase diagnostics measured every $10$ ns for three uniform drive-qubit detunings, $\Delta_g/2\pi=-50,0,+50$ kHz. Left panels show the normalized trace estimated from the diagonal walker normalization. Right panels show the phase angle in units of $10^{-4}$ rad. Error bars and shaded regions are obtained from bootstrap resampling.}
    \label{fig:qmc_trace_phase}
\end{figure}

\subsection{QMCtwin}

For a $97$-qubit open-system simulation, direct verification against exact dense master-equation evolution is not feasible, as the density matrix lives in a Liouville space of dimension $4^{97}\approx 2.5\times 10^{58}$. 
We therefore first examine internal consistency diagnostics of QMCtwin. 
Because the underlying Liouvillian is trace preserving, exact evolution must satisfy $\Tr[\rho(t)]=1$ throughout the simulation. In the QMC representation, summarized in Appendix~\ref{appendix-QMC} together with the relevant notation, we monitor the sample-averaged diagonal normalization
\beq
    \hat N_{\rm diag}(t)
    =
    \frac{1}{n_{\rm samp}}
    \sum_{r=1}^{n_{\rm samp}}N_{\rm diag}^{(r)}(t).
\eeq
The corresponding normalized trace diagnostic is
\beq
    \Tr[\rho^{\rm QMC}(t)]
    =
    \frac{\mathrm{Re}[\hat N_{\rm diag}(t)]}
    {N_{\rm diag}},
\eeq
while the associated phase,
\beq
    \phi(t) = \arctan\left[
    \frac{\mathrm{Im}[\hat N_{\rm diag}(t)]}
    {\mathrm{Re}[\hat N_{\rm diag}(t)]} \right],
\eeq
diagnoses residual complex phase accumulation in the walker normalization. Ideally, $\Tr[\rho^{\rm QMC}(t)]=1$ and $\phi(t)=0$.
As shown in Ref.~\cite{shen2025real}, under the population-dynamics assumptions analyzed there, the QMC estimator is unbiased and ergodic and converges to the exact master-equation solution. For a single independent population sample, the mean-square error decreases as  $O(1/N_{\rm diag})$, while averaging over $n_{\rm samp}$ independent samples gives a corresponding additional $1/n_{\rm samp}$ variance reduction.

In QMC, a sign or phase problem occurs when positive, negative, or complex-weight contributions nearly cancel, so that the desired signal becomes small compared with its sampling fluctuations~\cite{loh1990sign,troyer2005computational,mak1999multilevel,cohen2015taming,Marvian:2019aa}. In our signed-walker real-time estimator, this would appear as large fluctuations or complex phase accumulation in the diagonal normalization; stability of the sample-averaged diagonal normalization and its small phase therefore provide the corresponding practical diagnostic that sign-problem suppression remains effective. \cref{fig:qmc_trace_phase} shows these diagnostics, sampled every $10{\rm ns}$, for three representative detunings across the sweep. In all cases, the normalized trace stays within about $5\times10^{-3}$ of unity and the phase remains small ($\sim 10^{-4}$) throughout the $300$ ns circuit.
The residual trace drift should nevertheless be included in the systematic error budget for observables whose scale is $O(10^{-2})$.
Although these diagnostics are not an exact benchmark of all observables at this system size, they provide a useful self-consistency check that the QMC estimator remains approximately trace-preserving and sign-problem controlled before extracting syndrome and string-parity observables.

\subsection{Syndrome-to-string-parity information from QMCtwin}

Standard decoders are usually built around DEMs or sampled syndrome data derived from stochastic Pauli circuits~\cite{gidney2021stim,Higgott2023sparse}, whereas the QMC simulation estimates observables from a stochastic density-matrix representation from which coherent and correlated outcome distributions must first be extracted or compressed into decoder-facing data. Developing such a stochastic-density-matrix-informed decoding workflow is ongoing work. Here, in lieu of implementing a full decoder, we develop three complementary diagnostics to compare the information contained in the QMCtwin estimator with that retained by the Pauli-twirled Clifford model. 

The first two diagnostics test how faithfully individual stabilizer values are transferred to their measurement ancillas, and whether the resulting mismatch is directionally biased. More specifically, the syndrome-extraction bias $\delta_k$ of check operator $k$ measures the mismatch between the readout of ancilla $a(k)$ and the corresponding data stabilizer $S_k$, while the disagreement probability $p_k^{\neq}$ measures the probability that the corresponding two signs disagree. These two diagnostics are complementary.

The third diagnostic is more directly decoder-facing: it is a mutual information that quantifies how much a local set of measured $Z$-syndrome bits reduces uncertainty about a representative logical-$\bar Z$ string-parity proxy for $X$-type errors. This diagnostic depends on the full joint distribution of measured syndrome patterns and the string-parity proxies, as opposed to local checks. 

Together, these diagnostics distinguish local extraction bias, total ancilla-stabilizer disagreement, and syndrome-to-string-parity information available under realistic open-system dynamics. We now explain each diagnostic in detail.

\subsubsection{Syndrome-extraction bias $\delta_k$}
The first diagnostic measures the first-moment mismatch between each ancilla readout and the corresponding data stabilizer. In an ideal syndrome-extraction circuit, the ancilla outcome reports the eigenvalue of the corresponding data-qubit stabilizer, so the ancilla readout value and the data stabilizer have exactly the same expectation. Noise can break this equality of expectations through ancilla faults, data faults, or entangling-gate faults.

To quantify this effect, let $a(k)$ denote the ancilla qubit associated with check $k$. Using \cref{eq:toggling_frame_expect}, we evaluate the ancilla readout expectation in the toggling frame as
\beq\bal
    \expval{Z_{a(k)}}
    &=
    \Tr\left[
    \tilde{Z}_{a(k)}\tilde{\rho}(t_f)
    \right]\\
    \tilde Z_{a(k)}
    &=
    U_{\rm ideal}^{\dagger}(t_f)Z_{a(k)}U_{\rm ideal}(t_f),
\eal\eeq
where $t_f$ denotes the end of the circuit. Similarly, the corresponding data-stabilizer expectation is evaluated as
\beq
    \expval{S_k}
    =
    \Tr\left[
    \tilde S_k\tilde{\rho}(t_f)
    \right],
    \qquad
    \tilde S_k
    =
    U_{\rm ideal}^{\dagger}(t_f)S_k U_{\rm ideal}(t_f).
\eeq
Using the data stabilizer $S_k$ defined in \cref{eq:parity_check}, we define the syndrome-extraction bias as
\beq
\label{eq:delta_k}
    \delta_k := \expval{Z_{a(k)}} - \expval{S_k}.
\eeq
Thus, $\delta_k=0$ when the ancilla and data-stabilizer first moments agree, while nonzero $\delta_k$ indicates a systematic first-moment shift between the syndrome reported by the ancilla and the stabilizer value present on the data qubits at $t_f$.

\begin{figure}
    \centering
    \subfigure{\includegraphics[width=1.0\linewidth]{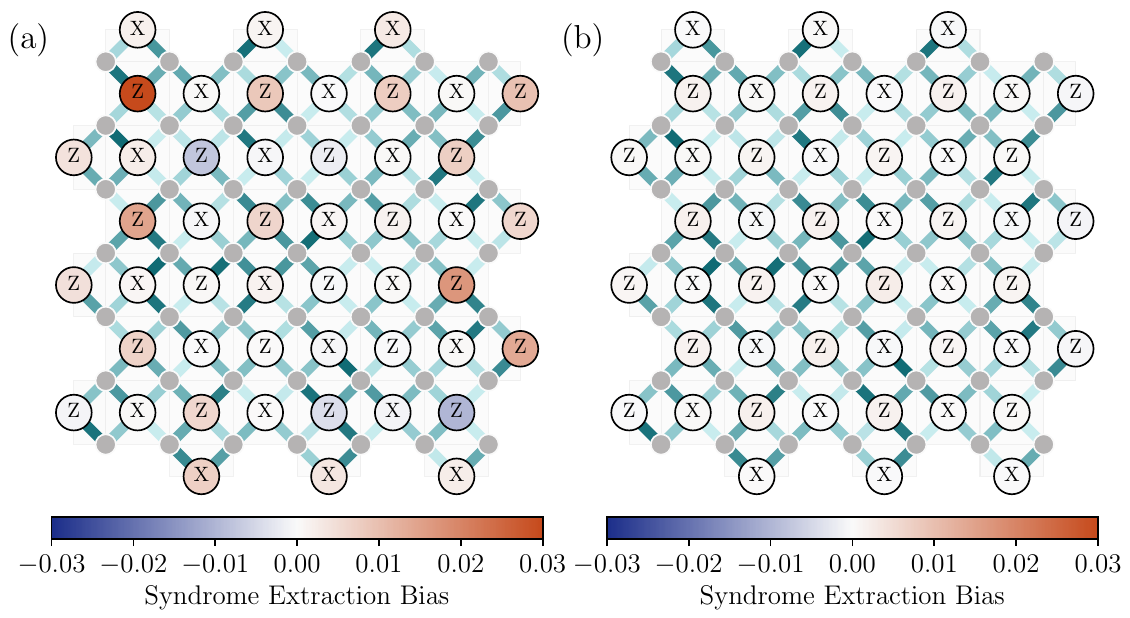}}
    \subfigure{\includegraphics[width=1.0\linewidth]{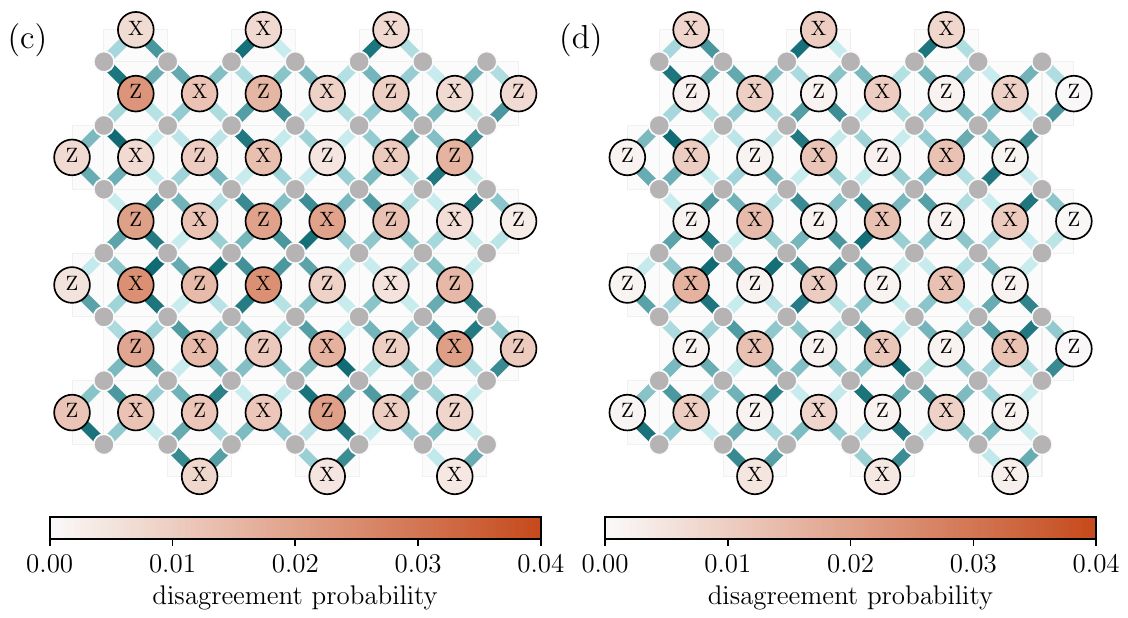}}
    \caption{Spatial distribution of (a),(b) syndrome-extraction bias and (c),(d) disagreement probability for the distance-$7$ rotated surface code at $\Delta_g=0$. For (a), (b), each labeled measurement qubit is colored by the bias of its check. (a),(c) Realistic noise QMC simulation. (b),(d) Clifford simulation using the noise parameters in \cref{tab:noise_model}. For (c),(d),     
    each measurement qubit is colored by $p^{\ne}_k$, the probability that the final ancilla readout disagrees with the corresponding data-stabilizer value. For $X$-type checks, the mean value is $1.16(\pm 0.02)\times10^{-2}$ in the QMC simulation and $1.03\times10^{-2}$ in the Stim simulation. For $Z$-type checks, the QMC mean is $1.24(\pm 0.02)\times10^{-2}$, whereas the Stim mean is $1.65\times10^{-3}$, smaller by about a factor of $7.5$. The parenthetical QMC uncertainties combine the five-run bootstrap with the residual trace-drift budget of \cref{fig:qmc_trace_phase}.}
\label{fig:syndrome_extraction_bias-p_k-neq}
\end{figure}

\Cref{fig:syndrome_extraction_bias-p_k-neq}(a,b) compares this quantity at $\Delta_g=0$ for the QMC master-equation simulation and for the Pauli-twirled Clifford baseline.  
To make the comparison controlled, the Stim circuit uses the same code geometry, gate schedule, and per-location gate and idle durations as the master-equation simulation. Since this comparison is at $\Delta_g=0$, the Stim baseline is the same-parameter Pauli-twirled reference model described in \cref{sec:model}, with relaxation, pure dephasing, residual $ZZ$ evolution, and coherent under-rotations replaced by the Pauli-twirled channels listed in Appendix~\ref{app:Pauli-twirling}. 
The two simulations therefore differ in whether these error mechanisms are evolved as continuous open-system dynamics or represented as stochastic Pauli channels at specified circuit locations.
In the Pauli-twirled model, the stochastic error probabilities in \cref{tab:noise_model} are mostly at the $10^{-4}$-$10^{-3}$ level per circuit location. Although a signed first-moment bias and a per-location fault probability are not directly comparable, the QMCtwin results show that the phase-sensitive and correlated structure discarded by this stochastic representation can nevertheless produce $O(10^{-2})$ syndrome-extraction biases that exceed the QMC uncertainty ({$\sim\pm1.6\times10^{-4}$}, combining the five-run bootstrap with the residual trace-drift budget of \cref{fig:qmc_trace_phase}) for the larger features; features comparable to this uncertainty should be interpreted with that budget in mind. These biases appear as spatially structured positive and negative features across the surface-code lattice, whereas the Pauli-twirled Clifford simulation produces a nearly featureless pattern on the same scale. Although this is not a direct logical-error-rate comparison, it shows that realistic open-system noise dynamics can reshape the local syndrome information supplied to the decoder in a way that is not reproduced by the reference stochastic Pauli model.

\subsubsection{Disagreement probability $p_k^{\neq}$}

The second diagnostic is the total probability that the ancilla readout disagrees with the corresponding data stabilizer. To define this probability, let
$A_k := Z_{a(k)}$.
Using the syndrome-label convention of \cref{eq:readout_stabilizer_relation}, the noiseless ideal circuit makes the outcome of $A_k$ agree with the eigenvalue of $S_k$. Since both $A_k$ and $S_k$ are $\pm 1$-valued observables, let
\beq
    p_{as}^{(k)}
    :=
    \Pr(A_k=a,S_k=s),
    \quad a,s\in\{\pm 1\} .
\eeq
Then the disagreement probability is
\beq
    p_k^{\neq}:=
    \Pr(A_k\neq S_k)    =
    p_{+-}^{(k)}+p_{-+}^{(k)}.
    \label{eq:pneq_joint_probs}
\eeq
The joint correlator is the expectation value 
\beq\bal
    \expval{A_k S_k} &= \sum_{a,s\in\{\pm 1\}}a s p_{as}^{(k)} \\
    &=
    p_{++}^{(k)}
    +
    p_{--}^{(k)}
    -
    p_{+-}^{(k)}
    -
    p_{-+}^{(k)}.
\eal\eeq
Using normalization $\sum_{a,s\in\{\pm 1\}}p_{as}^{(k)}=1$,
we obtain
\beq
    \expval{A_k S_k} = 1 - 2\left(p_{+-}^{(k)}+p_{-+}^{(k)}\right)
    =
    1-2p_k^{\neq}.
\eeq
Rearranging gives
\beq
    p_k^{\neq}
    =
    \frac{1-\expval{A_k S_k}}{2} = 
   \frac{1-\expval{Z_{a(k)}S_k}}{2}.
    \label{eq:disagreement_probability}
\eeq

In the toggling-frame simulation, the joint correlator entering \cref{eq:disagreement_probability} is evaluated as
\beq
    C_k :=
    \expval{Z_{a(k)}S_k}
    =
    \Tr\left[
        \widetilde{Z_{a(k)}S_k}(t_f)
        \tilde\rho(t_f)
    \right],
\eeq
where
\beq
    \widetilde{Z_{a(k)}S_k}(t_f)
    =
    U_{\rm ideal}^{\dagger}(t_f)
    Z_{a(k)}S_k
    U_{\rm ideal}(t_f).
    \label{eq:joint_correlator_toggling}
\eeq
Therefore,
\beq
    p_k^{\neq}
    =
    \frac{1-C_k}{2}.
\eeq

Now note that since
\beq
    \expval{A_k} = \sum_{a,s\in\{\pm 1\}} a p_{as}^{(k)} =
    p_{++}^{(k)}
    +
    p_{+-}^{(k)}
    -
    p_{-+}^{(k)}
    -
    p_{--}^{(k)},
\eeq
and, similarly,
\beq
    \expval{S_k} = \sum_{a,s\in\{\pm 1\}} s p_{as}^{(k)} 
    =
    p_{++}^{(k)}
    -
    p_{+-}^{(k)}
    +
    p_{-+}^{(k)}
    -
    p_{--}^{(k)},
\eeq
we have, using \cref{eq:delta_k},
\beq
    \delta_k
    =
    \expval{A_k}-\expval{S_k}
    =
    2\left(
        p_{+-}^{(k)}
        -
        p_{-+}^{(k)}\right).
    \label{eq:delta_different}
\eeq
Contrasting \cref{eq:pneq_joint_probs} with \cref{eq:delta_different}, we see that the two are complementary diagnostics as the sum and twice the difference of the two directional disagreement probabilities. In particular, $\delta_k$ can vanish even when $p_k^{\neq}$ is nonzero. Conversely, nonzero $\delta_k$ indicates that the disagreement is directionally biased. 

Finally, there is an equivalent, more direct way to evaluate $p_k^{\neq}$ in the ideal-circuit toggling frame. The toggling frame removes the ideal syndrome-extraction circuit, so the final ancilla Pauli $A_k=Z_{a(k)}$ in this frame represents the relative parity between the actual ancilla readout and the stabilizer value that would be reported by the ideal circuit. Thus the event $A_k=-1$ in the toggling frame is precisely the event that the ancilla readout disagrees with the corresponding data stabilizer. Therefore
\beq\bal
    p_k^{\neq}
    &=
    \Pr_{\tilde\rho(t_f)}(A_k=-1)
    =
    \Tr\left[
        \frac{I-A_k}{2}\tilde\rho(t_f)
    \right]\\
    &=
    \frac{1-\Tr[Z_{a(k)}\tilde\rho(t_f)]}{2}.
    \label{eq:pneq_toggling_direct}
\eal\eeq
We give a rigorous proof of this claim in Appendix~\ref{app:p_k-neq}. In practice, 
for the QMC estimator we replace $\Tr[Z_{a(k)}\tilde\rho(t_f)]$ with $\widehat{\Tr[Z_{a(k)}\tilde\rho(t_f)]}/\widehat{\Tr[\tilde\rho(t_f)]}$; see Appendix~\ref{appendix-QMC} for more details. For the Stim baseline, we estimate the same final-time event shot by shot: 
for each check $k$, the sampled final ancilla readout is compared with the tracked value of the corresponding data stabilizer $S_k$ at $t_f$, and $p_k^{\neq}$ is the empirical frequency with which the two signs disagree.

The resulting surface-code diagnostic is displayed in \cref{fig:syndrome_extraction_bias-p_k-neq}(c,d), which shows qualitatively different behavior in the QMC and Stim simulations for $X$- and $Z$-type checks. For $X$-type checks, the QMC and Stim disagreement probabilities are comparable. For $Z$-type checks, however, the QMC values remain on the same scale as the $X$-check values, while the Pauli-twirled Clifford baseline is lower by about a factor of 7.5.

This asymmetry can be understood from the different roles played by data and ancilla errors in the two check circuits. For a $Z$-type check, the data qubits are never basis-rotated: the circuit is effectively an interferometric measurement in which the measurement ancilla acquires a phase conditioned on a data $Z$ parity and is then rotated back for computational-basis readout. Since the input state is a $Z$-basis product state, many data-qubit $Z$-type or residual-$ZZ$ Pauli faults in the Stim model commute with the measured $Z$ stabilizers and do not, by themselves, change the final data parity. The remaining disagreement in the Pauli-twirled model is therefore produced mainly by stochastic faults that directly corrupt the ancilla parity readout, together with faults at circuit locations from which they propagate into ancilla-readout errors. In the master-equation simulation, by contrast, the ancilla readout is affected continuously during the check by coherent residual phases, pulse miscalibration, relaxation, and dephasing. These processes need not appear as discrete final data-parity flips; they can instead bias or shrink the ancilla signal relative to the final data stabilizer. Pauli twirling removes the coherent accumulation of these small phase shifts and the nonunital drift associated with relaxation, so the location-wise stochastic Clifford model can substantially underestimate the $Z$-check ancilla-stabilizer disagreement.

For $X$-type checks, the required Hadamard basis changes on both data and measurement qubits convert more of the same physical mechanisms into Pauli faults that affect the extracted check value in the stochastic model. The QMC and Stim disagreement probabilities are therefore closer in the $X$-check sector, while the largest discrepancy appears for $Z$ checks, where the master-equation dynamics predict an ancilla-readout mismatch that is mostly absent from the Pauli-twirled baseline.

Taken together with the signed syndrome-extraction bias in \cref{fig:syndrome_extraction_bias-p_k-neq}, these results show that the master-equation simulation captures error-diagnostic information that is absent or strongly suppressed in the Pauli-twirled Clifford baseline.

\subsubsection{Mutual information}

The diagnostics in the previous subsections quantify local check-extraction behavior. As our third diagnostic, we introduce a more decoder-facing information-theoretic measure. Rather than asking whether a single ancilla faithfully reports a single stabilizer, we ask how much information a local measured syndrome pattern carries about a non-local string-parity proxy.

Since the simulation is initialized in $\ket{0\cdots 0}$ and we consider a single syndrome-extraction round, we focus on $X$-type errors. Such errors are detected by $Z$-type stabilizer checks. They can also be probed by logical-$\bar Z$ string parities, since an $X$-type error anticommutes with any $Z$ string whose support it overlaps an odd number of times. The question we address here is how much the measured $Z$-syndrome reduces uncertainty about representative logical-$\bar Z$ string parities. We quantify this using mutual information, defined here so as to yield the average reduction in uncertainty about the string-parity variable after observing the syndrome information.

\begin{figure}[t]
    \centering
    \includegraphics[width=1.0\linewidth]{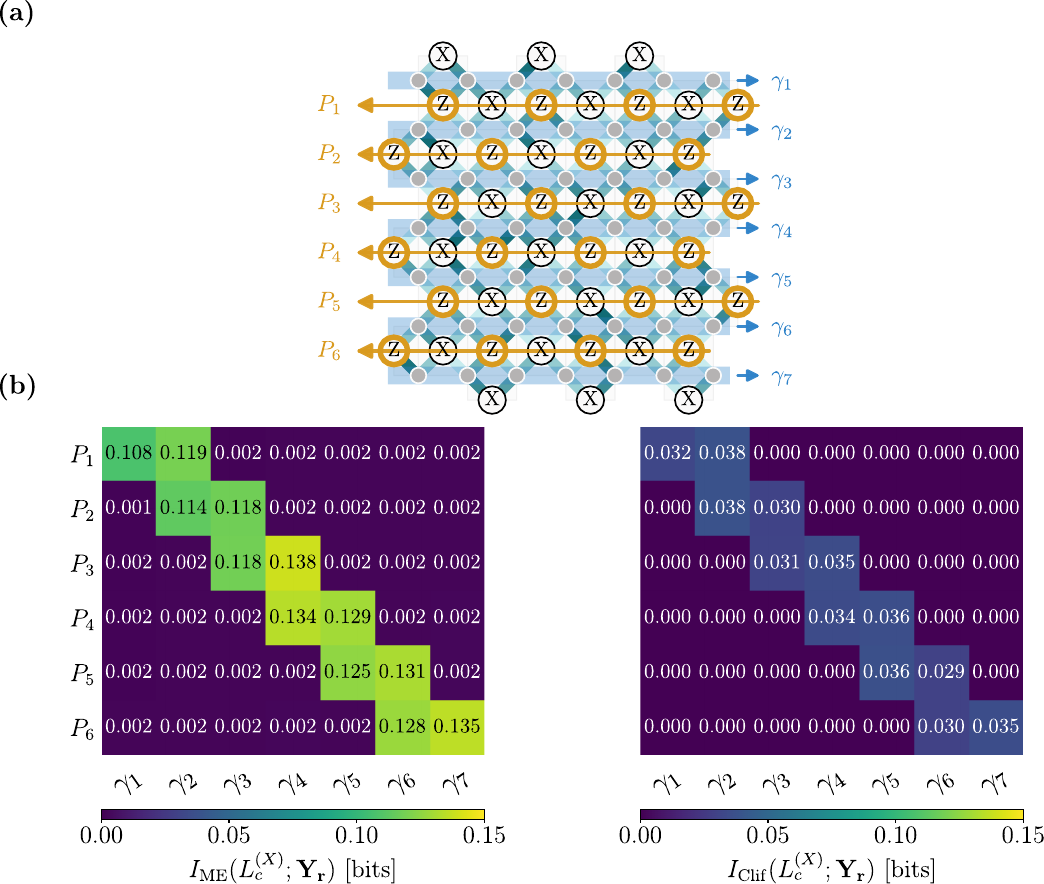}
    \caption{Mutual information between local row patches of $Z$-type syndrome checks and representative logical-$\bar Z$ string-parity proxies at $\Delta_g=0$. (a) Schematic of the analysis. The six row patches $P_1,\ldots,P_6$ each contain four $Z$-type syndrome checks, and the seven horizontal data-string supports $\gamma_1,\ldots,\gamma_7$ define minimum-weight representatives $\bar Z_{\gamma_c}$ of the logical-$\bar Z$ operator.
    (b) Mutual information $I_\mu(L^{(X)}_c ; \mathbf Y_r)$ between the four-bit row-patch syndrome variable $\mathbf Y_r$ of row patch $P_r$ and the corresponding $-1$ horizontal string-parity variable $L^{(X)}_c$. Values are computed from the QMC master-equation estimator (left) and from Stim samples using the Pauli-twirled stochastic noise model (right). The typical bootstrap standard error of the QMC mutual-information entries is ${\sim 0.0012}$ bits, estimated over the five independent QMC runs. The off-band entries are at or near this uncertainty scale, whereas the near-bidiagonal band entries are larger by about $0.1$ bits, supporting the observed near-bidiagonal structure.}
    \label{fig:entropy_reduction}
\end{figure}

We begin with the string-parity proxy.
As illustrated in \cref{fig:entropy_reduction}(a), we choose the seven horizontal data-qubit-string supports in the distance-7 rotated surface-code layout as canonical minimum-weight representatives of the logical-$\bar Z$ operator. Let
\beq
    \Gamma_{\bar Z}
    :=
    \{\gamma_1,\ldots,\gamma_7\}
\eeq
denote these seven supports, where $\gamma_c\subset\mathcal D$ is the set of seven data qubits in horizontal data string $c$. The corresponding logical-$\bar Z$ representative is
\beq
    \bar Z_{\gamma_c}
    =
    \prod_{q\in\gamma_c} Z_q .
\eeq

The operators $\bar Z_{\gamma_1},\ldots,\bar Z_{\gamma_7}$ are not independent logical operators: in the codespace, they are stabilizer-equivalent representatives of the same logical-$\bar Z$ operator. In the noisy single-round circuit studied here, however, local $X$-type faults can flip the parity of some horizontal string representatives but not others. The accompanying $Z$-type syndrome checks can therefore carry local information about which representative string parities have flipped. We use these horizontal string parities as probes of $X$-type error structure.

For each support $\gamma_c\in\Gamma_{\bar Z}$, define the projector onto the $-1$ parity sector of the corresponding $Z$ string:
\beq
    L_c^{(X)} := L_{\gamma_c}^{(X)} := \frac{I-\bar Z_{\gamma_c}}{2}.
\eeq
For the initial product state, $\bar Z_{\gamma_c}=+1$, so the outcome $L_c^{(X)}=1$ flags a flip of the chosen horizontal string representative. We use the same symbol $L_c^{(X)}$ for the projector and for the associated binary outcome variable. In a fully decoded code state, an uncorrected nontrivial $X$ chain crossing such a representative would correspond to a logical-$X$ fault. In the single-round diagnostics used here, however, $L_c^{(X)}$ is a logical-string-parity proxy for $X$-type error structure, not a decoded logical-failure label.

Having defined the string-parity proxy, we next consider the syndrome associated with $Z$-type stabilizer checks. 
The distance-$7$ rotated code has $24$ $Z$-type checks [the $Z$-labeled circles in \cref{fig:surface_code}(a)]. Enumerating the full $Z$-syndrome distribution would therefore require $2^{24}$ outcomes. Instead, as shown in \cref{fig:entropy_reduction}(a), we coarse-grain the $Z$-type syndrome by partitioning these checks into six row patches $\{P_r\}_{r=1}^6$, where each row patch $P_r$ contains four $Z$-type checks. 
For this analysis, after restricting to the $24$ $Z$-type checks, we relabel them row by row as
$\mathcal C_Z=\{1,\ldots,24\}$.
The six row patches are defined by
\beq
    P_r
    =
    \{k_{r1},k_{r2},k_{r3},k_{r4}\},
    \qquad
    r=1,\ldots,6,
\eeq
where
\beq
    k_{r\alpha}
    =
    4(r-1)+\alpha,
    \qquad
    \alpha=1,\ldots,4.
\eeq
Thus $P_r$ is the set of four integer labels corresponding to the $Z$-type checks in row $r$.

Let
\beq
     \mathbf Y_r
    =
    (Y_{r1},Y_{r2},Y_{r3},Y_{r4})
    \in\{0,1\}^4
\eeq
denote the measured syndrome-bit vector for row patch $P_r$, where $Y_{r\alpha}=0$ corresponds to a $+1$ readout of check $k_{r\alpha}$ and $Y_{r\alpha}=1$ corresponds to a $-1$ readout. 
We write
\beq
    \mathbf y
    =
    (y_{1},y_{2},y_{3},y_{4})
    \in\{0,1\}^4
\eeq
for a particular syndrome pattern. 
Rewriting \cref{eq:readout_projector}, the corresponding marginal final readout projector for row patch $P_r$ in the rotating control frame is
\beq
    M_{r,\mathbf y}^{\rm rd}
    =
    \prod_{\alpha=1}^{4}\frac{I+(-1)^{y_{\alpha}}Z_{a(k_{r\alpha})}}{2}.
    \label{eq:row_patch_projector}
\eeq

For a fixed noise model $\mu\in\{{\rm ME},{\rm Clif}\}$, define the conditional probability that the horizontal string $\gamma_c$ is in the flipped, $-1$ sector given a syndrome $\mathbf y$ from row-patch $r$:
\beq
    \pi^{\mu}_{c,r}(\mathbf y) =
    \Pr_{\mu}\left(L_c^{(X)}=1 \mid \mathbf Y_r=\mathbf y\right), \quad \Pr_\mu(\mathbf Y_r=\mathbf y)>0 .
\eeq
For syndrome patterns with $\Pr_\mu(\mathbf Y_r=\mathbf y)=0$, the value of $\pi^{\mu}_{c,r}(\mathbf y)$ has no effect on the entropy sums below.
The conditional entropy of the horizontal string-parity variable given the row-patch syndrome is
\beq
    H_{\mu}\left(L_c^{(X)}\mid \mathbf Y_r\right) =
     \sum_{\mathbf y\in\{0,1\}^4}
    \Pr_{\mu}(\mathbf Y_r=\mathbf y)
    h_2\left(
        \pi^{\mu}_{c,r}(\mathbf y)
    \right),
\eeq
where 
\beq
h_2(q)=-q\log_2 q-(1-q)\log_2(1-q)
\eeq
is the binary entropy. The marginal entropy is
\beq
    H_{\mu}\left(L_c^{(X)}\right) = h_2\left[\Pr_{\mu}\left(L_c^{(X)}=1\right)\right].
\eeq
The mutual information between the row-patch syndrome and the binary horizontal string-parity variable is
\beq
    I_{\mu}\left(L_c^{(X)} ; \mathbf Y_r\right)=
    H_{\mu}\left(L_c^{(X)}\right)-
    H_{\mu}\left(L_c^{(X)}\mid \mathbf Y_r\right).
\eeq
Larger $I_{\mu}$ means that the local row-patch syndrome removes more uncertainty about the corresponding string-parity proxy.

For the master-equation simulation, the required joint probabilities are evaluated directly from the QMC stochastic density-matrix estimator. 
Define
\beq
    P_{\ell,c}
    =
    \frac{I+(-1)^\ell\bar Z_{\gamma_c}}{2},\quad \ell\in\{0,1\},
\eeq
so that $\ell=1$ corresponds to the $-1$ string-parity sector. Then
\beq
    \Pr_{\rm ME}\left(L_c^{(X)}=\ell,\mathbf Y_r=\mathbf y\right)
    =
    \Tr\left[\widetilde{P_{\ell,c}M_{r,\mathbf y}^{\rm rd}}(t_f)
    \tilde\rho(t_f)
    \right],
\eeq
where 
\beq
    \widetilde{
        P_{\ell,c}
        M_{r,\mathbf y}^{\rm rd}
    }(t_f)
    =
    U_{\rm ideal}^{\dagger}(t_f)
    P_{\ell,c}
    M_{r,\mathbf y}^{\rm rd}
    U_{\rm ideal}(t_f).
\eeq
In the numerical evaluation, for each pair $(c,r)$ we form the $2\times 2^4$ joint distribution over $\ell\in\{0,1\}$ and $\mathbf y\in\{0,1\}^4$ and normalize it over this outcome set before computing entropies. We use the real parts of the QMC probability estimates after verifying that the imaginary parts are negligible on the scale of the plotted quantities. Any residual normalization drift is thereby treated as part of the QMC estimator error shown in \cref{fig:qmc_trace_phase}.
For the Clifford baseline, the same quantities are estimated from Stim samples generated using the Pauli-twirled stochastic noise model.

\begin{figure}[t]
    \centering
    \includegraphics[width=0.75\linewidth]{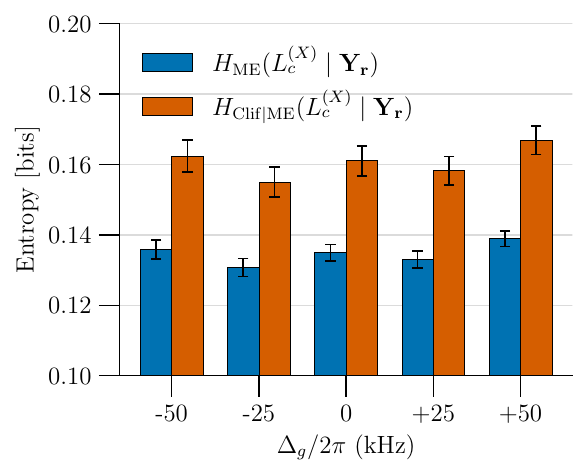}
    \caption{Conditional string-parity uncertainty for ME simulations at various detunings compared with the fixed Pauli-twirled Stim predictor defined in \cref{sec:model}. Blue bars show the ME conditional entropy $H_{\rm ME}(L_c^{(X)}\mid\mathbf Y_r)$, evaluated using ME row-patch syndrome weights and ME conditional string-parity probabilities. Orange bars show the cross entropy $H_{\rm Clif|ME}(L_c^{(X)}\mid\mathbf Y_r)$, obtained by using the ME conditional probabilities as the target distribution and the Stim conditional probabilities as the predictor. Bars are averaged over the same row-patch and horizontal-string pairs used in \cref{fig:entropy_reduction}, and error bars are obtained by bootstrap resampling.}
    \label{fig:conditional_entropy}
\end{figure}

\cref{fig:entropy_reduction}(b) shows $I_\mu(L_c^{(X)} ; \mathbf Y_r)$ for all six row patches and all seven horizontal string-parity variables at $\Delta_g=0$. Two features stand out: the mutual information is substantially larger in the master-equation simulation, and both simulations exhibit a bidiagonal structure.
Regarding the first feature, the dominant entries in the master-equation simulation are approximately $0.11$-$0.14$ bits, nearly four times larger than the corresponding Pauli-twirled Clifford values of $0.03$-$0.04$ bits. Thus, under its own joint distribution, the master-equation simulation predicts substantially stronger local syndrome-to-string-parity correlations than the stochastic Clifford model. Because each mutual information is evaluated under its own joint distribution, this comparison quantifies the syndrome information content of each model, not the cost of substituting one model for the other; we address the latter below, in the cross-entropy comparison in \cref{fig:conditional_entropy}.

To explain the second, bidiagonal feature, note first that, as shown in  \cref{fig:entropy_reduction}(a), the patch $P_r$ is the row of $Z$-type checks lying between the neighboring horizontal string representatives $\gamma_r$ and $\gamma_{r+1}$. Equivalently, with $S_k^Z$ as defined in \cref{eq:parity_check}, the product of the $Z$-type stabilizers in this patch cancels all interior data-qubit $Z$ factors and leaves the strip of two neighboring string representatives:
\beq
    \prod_{k\in P_r} S_k^Z
    =
    \bar Z_{\gamma_r}\bar Z_{\gamma_{r+1}} 
    \qquad
    r=1,\ldots,6,
    \label{eq:strip_identity}
\eeq
Thus, the mod-$2$ parity of the row-patch syndrome bits is tied to the relative string parity $L_r^{(X)}\oplus L_{r+1}^{(X)}$. If the two neighboring string-flip variables were independent unbiased bits, such relative-parity information alone would not give large mutual information with either one separately. The single-round distribution is far from that limit: string flips are sparse and local, and the full four-bit pattern $\mathbf Y_r$ (not only its parity), carries information about where in the strip an $X$-type fault occurred.
Thus, $\mathbf Y_r$ reduces uncertainty primarily about the two neighboring single-string variables $L_r^{(X)}$ and $L_{r+1}^{(X)}$, while correlations with nonadjacent string representatives require longer error chains, multiple faults, or long-range correlations. Because simultaneous flips of both neighboring representatives are rare after a single round, the relative-parity information carried by $\mathbf Y_r$ translates into nonzero mutual information with each of $\bar Z_{\gamma_r}$ and $\bar Z_{\gamma_{r+1}}$ individually. This explains the dominant entries near $(P_r,\gamma_r)$ and $(P_r,\gamma_{r+1})$ and the much smaller remaining entries.

The entropy-reduction comparison shows that the two models contain different amounts of syndrome information about the horizontal string-parity variables, but those quantities are computed under different joint distributions. To compare conditional predictions on the same reference distribution, we fix the master-equation distribution and evaluate the Pauli-twirled Clifford predictor on it. For each row-patch syndrome pattern $\mathbf y$, we evaluate the Stim conditional probability on the ME reference distribution using the binary cross entropy
\beq\bal
    H_{\rm Clif|ME}(L_c^{(X)}\mid \mathbf Y_r)
    &=
     \sum_{\mathbf y\in\{0,1\}^4}
    \Pr_{\rm ME}(\mathbf Y_r=\mathbf y)\times\\
    &\quad h_{\times}\left(
        \pi^{\rm ME}_{c,r}(\mathbf y),
        \tilde \pi^{\rm Clif}_{c,r}(\mathbf y)
    \right),
\eal\eeq
where 
\beq
h_{\times}(p,q)=-p\log_2 q-(1-p)\log_2(1-q)
\eeq 
is the binary cross entropy. In the finite-sample Stim estimate, 
$\tilde \pi^{\rm Clif}_{c,r}(\mathbf y)$ denotes the Jeffreys-smoothed
conditional probability: if $n_{\ell}^{(c,r)}(\mathbf y)$ is the number of Stim samples with row-patch syndrome $\mathbf Y_r=\mathbf y$ and $L_c^{(X)}=\ell$, then
\beq
    \tilde \pi^{\rm Clif}_{c,r}(\mathbf y)
    =
    \frac{n_1^{(c,r)}(\mathbf y)+1/2}{n_0^{(c,r)}(\mathbf y)+n_1^{(c,r)}(\mathbf y)+1}.
\eeq
The difference,
\beq\bal
    &D_{\mathrm{KL}}^{\mathrm{ME} \to \mathrm{Clif}}(L_c^{(X)} \mid \mathbf{Y}_r) \\
    &\quad = H_{\rm Clif|ME}(L_c^{(X)}\mid \mathbf Y_r) - H_{\rm ME}(L_c^{(X)}\mid \mathbf Y_r),
\eal\eeq
is the ME-weighted Kullback-Leibler (KL) divergence between the ME conditional distribution and the smoothed Clifford conditional predictor. 
In the infinite-sample limit with no smoothing bias, $D_{\mathrm{KL}}$ vanishes only when the Clifford model reproduces the ME conditional probabilities for every row-patch syndrome pattern with nonzero ME weight.

\cref{fig:conditional_entropy} shows the corresponding conditional entropies and cross entropies over the detuning sweep, averaged over the same row-patch and horizontal-string pairs used in \cref{fig:entropy_reduction}. For all detunings shown, the ME-referenced Clifford cross entropy is larger than the ME conditional entropy, so the averaged KL gap is positive. Since the two quantities are evaluated on the same ME row-patch syndrome distribution, this gap reflects a mismatch in the conditional probabilities of the $-1$ string-parity sector, not merely a difference in how often row-patch syndrome patterns occur.

Operationally, the KL gap is the excess log-loss, in bits, incurred by using the smoothed Clifford conditional predictor instead of the ME conditional predictor on data drawn from the ME reference distribution. Thus a positive gap indicates that the Pauli-twirled Clifford model does not reproduce conditional syndrome-to-string-parity structure present in the QMCtwin results. 

If the master-equation distribution estimated by QMCtwin is taken as the reference distribution for the simulated device state, then a decoder or predictor calibrated only to the Pauli-twirled Clifford model would assign different conditional probabilities to the string-parity variables, potentially leading to degraded decoding performance once embedded in a full decoding workflow. This provides a decoder-facing indication that the master-equation simulation contains syndrome-to-string-parity structure that is absent or strongly suppressed by the stochastic Pauli representation, and that could be useful once incorporated into a full decoding workflow.

\section{Discussion}
\label{sec:discussion}

We have demonstrated a QMC master-equation simulation at an experimentally relevant QEC circuit scale. A brute-force dense treatment of the full $97$-qubit open-system dynamics is completely infeasible, with a Liouville-space dimension $4^{97}$. Simulating the same microscopic master equation by uncompressed quantum trajectories would still require evolving $2^{97}$-dimensional state vectors, while tensor-network approaches at this scale would likewise require  compression, truncation, or structure-specific approximations. The simulated scale is directly connected to present surface-code experiments. At this size, a microscopic open-system model can be used to estimate circuit-level syndrome statistics, rather than only isolated-gate characterizations. The QMC estimator contains syndrome correlations and syndrome-to-string-parity information that can be compressed into decoder-facing statistical models. This suggests a route to constructing detector-error models \textit{ab initio} from realistic outcome distributions, rather than deriving them solely from a Pauli stochastic approximation. Extending this procedure to full multi-round detector histories, logical observables, and ultimately the quantum instrument~\cite{hashim2025practical} associated with the realistic syndrome-extraction process remains future work. The same estimator can also be used to sample syndrome data labeled by chosen string-parity or fault-proxy observables for training more expressive neural-network decoders~\cite{bausch2024learning}. 

This capability can also be used to compare QEC architectures under realistic hardware noise, such as quantum LDPC codes where nonlocal check structure may interact strongly with correlated errors. Beyond QEC, the ability to simulate realistic many-qubit open-system circuits provides a path toward optimizing pulse schedules~\cite{magann2021pulses,dong2021doubly,kosut2025fundamental}, dynamical-decoupling protocols~\cite{tripathi2025qudit,vezvaee2026demonstration}, and device layout at the level of complete algorithms or error-correction cycles~\cite{mathews2026placing}. Thus, QMC master-equation simulation provides a route from microscopic hardware models to decoder- and control-relevant circuit statistics, allowing realistic open-system simulations to inform the design of QEC architectures and hardware-control protocols.

\section{Data Availability}
All scripts and data needed to reproduce the results in this work are available online~\cite{data_repo}.

\section{Acknowledgements}
The authors gratefully acknowledge Amazon Web Services (AWS) for providing cloud computing resources that supported the numerical simulations reported in this work. This material is based upon work supported by, or in part by, the U. S. Army Research Laboratory and the U.S. Army Research Office under contract/grant number W911NF2310255.

\appendix

\section{Transformation from the lab frame to the rotating control frame}
\label{app:RF}

Here we explain how we transform \cref{eq:lindblad_lab} to the rotating control frame. Let $\omega_i^{\rm ref}(t)$ be the reference angular frequency used to define the microwave control frame for qubit $i$. We take the transformation to the rotating control frame to be the product of local $Z$ rotations
\beq
    R(t)
    =
    \exp \left[
        \frac{i}{2}
        \sum_i
        \theta_i(t)\sigma_i^z
    \right],
    \quad
    \theta_i(t)
    =
    \int_0^t \omega_i^{\rm ref}(t') dt' .
\eeq
With the lab-frame convention
$H_{\rm lab} = -\frac{1}{2}\sum_i\omega_{q,i}\sigma_i^z$,
this transformation converts the large bare precession into the residual rotating-frame detuning
$-\frac{1}{2}\sum_i \bigl(\omega_{q,i}-\dot\theta_i(t)\bigr)\sigma_i^z$.
Thus, choosing $\dot\theta_i=\omega_{q,i}$ removes the bare qubit precession, while choosing a different reference frequency leaves only the corresponding residual detuning. The Lindblad operators transform as $L_\ell(t)=R^\dagger(t)L_\ell^{\rm lab}R(t)$, leaving the dissipator in the form used in \cref{eq:lindblad_control}. Virtual-$Z$ updates may equivalently be represented as changes of the control-frame phases or as updates to the phases of subsequent pulses; in our simulations we use the latter convention and include these frame updates in the pulse phases $\phi_g$, not as additional physical Hamiltonian terms.

\section{QMC algorithm}
\label{appendix-QMC}

Here we summarize the QMC algorithm developed in Ref.~\cite{shen2025real}, adapting it to the surface code simulations described in this work.

After column-vectorizing the toggling-frame density matrix, the master equation becomes a linear ordinary differential equation in Liouville space,
\beq
    \kket{\dot{\tilde{\rho}}(t)} = \tilde{\mathcal L}(t)\kket{\tilde{\rho}(t)},
    \label{eq:ode_vectorized}
\eeq
where $\tilde{\mathcal L}(t)$ is the Liouvillian corresponding to \cref{eq:lindblad_toggling_general},
\begin{align}
    \tilde{\mathcal L}(t)
    =&-i\left(I\otimes \tilde H(t)-\tilde H^{T}(t)\otimes I\right)
    \nonumber\\
    &+\sum_{\ell}\gamma_{\ell}(t)\Big[
        \tilde L_{\ell}^{*}(t)\otimes \tilde L_{\ell}(t)
        -\frac12 I\otimes \tilde L_{\ell}^{\dagger}(t)\tilde L_{\ell}(t) \nonumber\\
    &-\frac12 \big(\tilde L_{\ell}^{\dagger}(t)\tilde L_{\ell}(t)\big)^T\otimes I
    \Big].
    \label{eq:liouvillian_vectorized}
\end{align}
For $n$ qubits, the Hilbert-space dimension is $D=2^n$, so the vectorized density matrix $\kket{\tilde{\rho}(t)}$ lives in a $D^2$-dimensional Liouville space. Rather than storing all $D^2$ amplitudes explicitly, QMC represents each sample $r$ by a sparse complex-integer-valued population vector
\beq
    \kket{\tilde N^{(r)}(t)}
    = 
    \sum_{\alpha=1}^{N_{\rm tot}^{(r)}(t)}s_{\alpha}^{(r)}(t) \kket{i_{\alpha}^{(r)},j_{\alpha}^{(r)}},
    \label{eq:walker_representation}
\eeq
where $\kket{i,j}={\rm vec}(\ketb{i}{j})$ is a Liouville-space basis element, $s_{\alpha}^{(r)}(t)\in\{\pm 1,\pm i\}$,  and a walker is the signed basis contribution
\beq
    w_{\alpha}^{(r)}(t)
    =
    s_{\alpha}^{(r)}(t)
    \kket{i_{\alpha}^{(r)},j_{\alpha}^{(r)}} .
\eeq 
Thus, a walker $w_{\alpha}^{(r)}(t)$ is a single unit of complex-signed population at a Liouville-space basis location; multiple walkers may occupy the same location, and walkers with signs $+1$ and $-1$, or $+i$ and $-i$, annihilate upon coalescence. 
The toggling-frame density matrix is estimated by ensemble averaging with the fixed initial normalization,
\beq
    \kket{\hat{\tilde{\rho}}(t)}
    =
    \frac{1}{n_{\rm samp}N_{\rm diag}}
    \sum_{r=1}^{n_{\rm samp}}
    \kket{\tilde N^{(r)}(t)} ,
 \eeq
where $n_{\rm samp}$ is the number of samples.
The trace of this finite-QMC estimator is determined by the sample-averaged diagonal population,
\beq
 \begin{aligned}
   \hat N_{\rm diag}(t)
    &=
    \frac{1}{n_{\rm samp}}
    \sum_{r=1}^{n_{\rm samp}}N_{\rm diag}^{(r)}(t),\\
    N_{\rm diag}^{(r)}(t)
    &=
    \sum_{\alpha:i_{\alpha}^{(r)}=j_{\alpha}^{(r)}}
    s_{\alpha}^{(r)}(t).
\end{aligned}
\eeq
Equivalently,
\beq
    \widehat T(t)
    :=
    \widehat{\Tr[\tilde\rho(t)]}
    =
    \frac{\hat N_{\rm diag}(t)}{N_{\rm diag}} .
\eeq
For expectation values and probabilities reported in the main text, the exact quantities are defined with respect to a normalized density matrix. In the QMC estimator, we therefore use the trace-normalized estimator
\beq
    \widehat{\expval{O}}(t) = \frac{\widehat{\Tr[O\tilde\rho(t)]}}{\widehat{\Tr[\tilde\rho(t)]}},
\eeq
unless explicitly stated otherwise. The residual deviation of $\widehat T(t)$ from unity is therefore both the normalization correction entering reported observables and the trace-preservation diagnostic used in the main text.

The simulation is initialized with $N_{\rm diag}^{(r)}(0)=N_{\rm diag}$ for each sample. Since the Liouvillian is trace preserving, $\mathrm{Re}[\hat N_{\rm diag}(t)]$ should remain close to $N_{\rm diag}$, with a negligible imaginary component, when sign suppression is effective. Thus, $N_{\rm diag}$ serves as the main accuracy-cost control parameter: larger $N_{\rm diag}$ yields smaller statistical error and higher memory/runtime overhead. Because QMC stores and updates only occupied locations, the cost of each step is governed by the instantaneous support of $\kket{\tilde N^{(r)}(t)}$ and the local sparsity of $\tilde{\mathcal L}(t)$, rather than by the full $D^2$ Liouville dimension. In the noisy-circuit setting of interest here, relaxation and pure dephasing suppress 
large portions of the coherent off-diagonal support, so the number of occupied Liouville-space locations sampled by QMC can remain much smaller than the full $D^2$ dimension. For the regimes studied in Ref.~\cite{shen2025real}, the effective sampled support was observed to scale as $O(\lambda D)$ with $\lambda\ll1$; see that reference
for the detailed scaling argument and numerical evidence. 

\cref{eq:ode_vectorized} is integrated with the adaptive, variable-coefficient Adams-Bashforth-Moulton (VCABM) solver~\cite{rackauckas2017differentialequations}.
In a linear multistep VCABM method, the time integral of the derivative $\tilde{\mathcal L}(t)\kket{\tilde{\rho}(t)}$ over each step is approximated by linear combinations of $\tilde{\mathcal L}(t_m)\kket{\tilde{\rho}(t_m)}$ evaluated at previous and corrected times, up to fifth order. 
In the QMC implementation, each 
required Liouvillian-vector product is estimated by stochastic population dynamics. 
For a fixed time step and a fixed current population, the spawning step gives an unbiased estimator of the Liouvillian increment,
\beq
    \mathbb E\left[
        \kket{\Delta \tilde N^{(r)}(t)}
        \middle|
        \kket{\tilde N^{(r)}(t)}
    \right]
    =
    \Delta t\tilde{\mathcal L}(t)\kket{\tilde N^{(r)}(t)}.
\label{eq:spawn_unbiased}
\eeq
If this increment is used as a single first-order Euler step for the differential equation, the corresponding local time-discretization error is $O(\Delta t^2)$. In the simulations, the VCABM step combines   stochastic derivative estimates of this form with the corresponding multistep coefficients; convergence of the resulting population-dynamics estimator is discussed in Ref.~\cite{shen2025real}.
Specifically, let
\beq
    \tilde{\mathcal L}^{kl}_{ij}(t) = \bbra{k,l} \tilde{\mathcal L}(t) \kket{i,j} .
\eeq
For a walker $w_{\alpha}^{(r)}(t)=s_{\alpha}^{(r)}(t)\kket{i,j}$, define the per-walker expected spawn count over a time step $\Delta t$ as
\beq
    \lambda_{ij}^{\rm sp}(t)
    =
    \Delta t
    \sum_{k,l}
    \left(
        \left|{\rm Re}[\tilde{\mathcal L}^{kl}_{ij}(t)]\right|
        +
        \left|{\rm Im}[\tilde{\mathcal L}^{kl}_{ij}(t)]\right|
    \right).
    \label{eq:spawn_prob}
\eeq
The fractional part of this expected count defines the Bernoulli probability for one additional stochastic spawn attempt,
\beq
    p_{ij}^{\rm sp}(t) = \lambda_{ij}^{\rm sp}(t) - \lfloor \lambda_{ij}^{\rm sp}(t)\rfloor . 
\eeq
The time-step $\Delta t$ is varied adaptively based on the $L_1$ norm between the Adams-Bashforth predictor increment and the Adams-Moulton corrector increment. Each walker at location $(i,j)$ produces $\lfloor \lambda_{ij}^{\rm sp}(t) \rfloor$ deterministic spawn attempts and one additional Bernoulli spawn attempt with probability $p_{ij}^{\rm sp}(t)$.
Equivalently, if $N_{ij}^{(r)}(t)$ walkers occupy location $(i,j)$ in sample $r$, the total number of spawned children from that location is distributed as
\beq
    \begin{aligned}
    N_{ij}^{\rm sp}(t)
    = \lfloor \lambda_{ij}^{\rm sp}(t) \rfloor N_{ij}^{(r)}(t) + N^{\rm stoch}_{ij}(t) \\
    N^{\rm stoch}_{ij}(t) \sim {\rm Binomial}
    \left[
        N_{ij}^{(r)}(t),
        p_{ij}^{\rm sp}(t)
    \right].
    \end{aligned}
\eeq
For each spawn attempt, whether deterministic or stochastic, the target channel $(k,l,c)$, with $c\in\{{\rm Re},{\rm Im}\}$, is sampled according to
\beq
    p_{ij\rightarrow(kl,c)}(t)
    =
    \frac{\left|c\left[\tilde{\mathcal L}^{kl}_{ij}(t)\right]\right|}
    {\sum_{k',l'}
    \left(
        \left|{\rm Re}[\tilde{\mathcal L}^{k'l'}_{ij}(t)]\right|
        +
        \left|{\rm Im}[\tilde{\mathcal L}^{k'l'}_{ij}(t)]\right|
    \right)}.
    \label{eq:spawn_channel_prob}
\eeq
If $\lambda_{ij}^{\rm sp}(t)=0$, no spawn attempt is generated from that walker.
Each spawned child at $\kket{k,l}$ carries sign
\beq
    s_{\alpha}^{\prime(r)}(t)
    =
    \begin{cases}
        s_{\alpha}^{(r)}(t){\rm sgn}\left({\rm Re}[\tilde{\mathcal L}^{kl}_{ij}(t)]\right),
        & c={\rm Re}, \\
        is_{\alpha}^{(r)}(t){\rm sgn}\left({\rm Im}[\tilde{\mathcal L}^{kl}_{ij}(t)]\right),
        & c={\rm Im}.
    \end{cases}
    \label{eq:spawn_sign}
\eeq
Summing all spawned walkers yields the stochastic increment in \cref{eq:spawn_unbiased}. After spawning, walkers on the same basis location are merged and opposite signs annihilate. This keeps the update cost proportional to the occupied support while continuously suppressing sign growth during the real-time evolution. Under the assumptions analyzed in Ref.~\cite{shen2025real}, the QMC estimator is unbiased and ergodic and converges to the exact master-equation solution. For a single independent population sample, the mean-square error decreases as $O(1/N_{\rm diag})$; averaging over $n_{\rm samp}$ independent samples gives the corresponding additional $1/n_{\rm samp}$ variance reduction.  Stability of the sample-averaged diagonal normalization $\hat N_{\rm diag}(t)$ is the practical diagnostic used here to verify that the simulation remains in the sign-problem-controlled regime.

\section{Additional Modeling Details}
\label{app:additional-modeling}

\begin{figure}[t]
    \centering
    \includegraphics[width=1.0\linewidth]{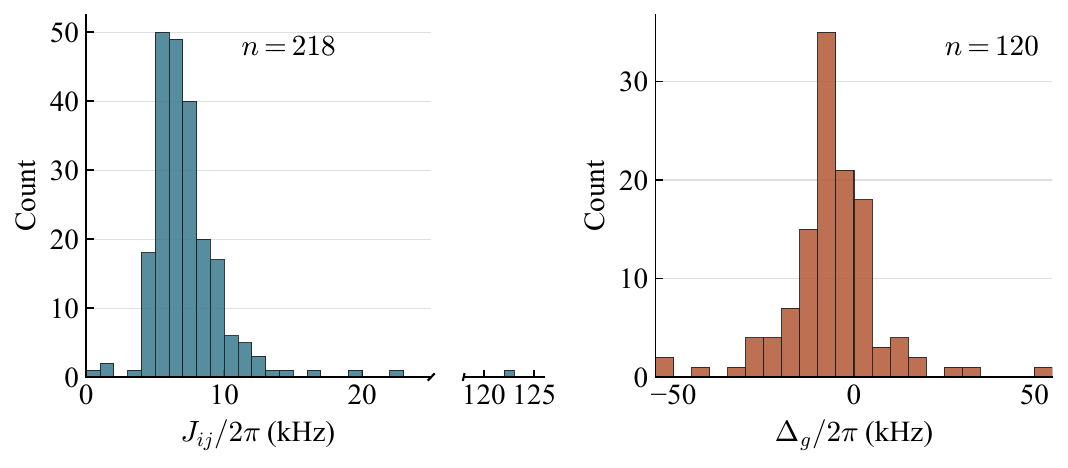}
    \caption{Distribution of coupler crosstalk strengths (left) and qubit detunings (right) on \texttt{ibm\_miami}. The parameters are extracted using diagnostic sequences applied to qubits in an order determined by the device graph. The histograms show characterization data used to motivate the representative simulation ranges in \cref{tab:noise_model}. The data were collected on January 13, 2026.}
    \label{fig:graph_spec}
\end{figure}

\begin{figure}[t]
    \centering
    \includegraphics[width=1.0\linewidth]{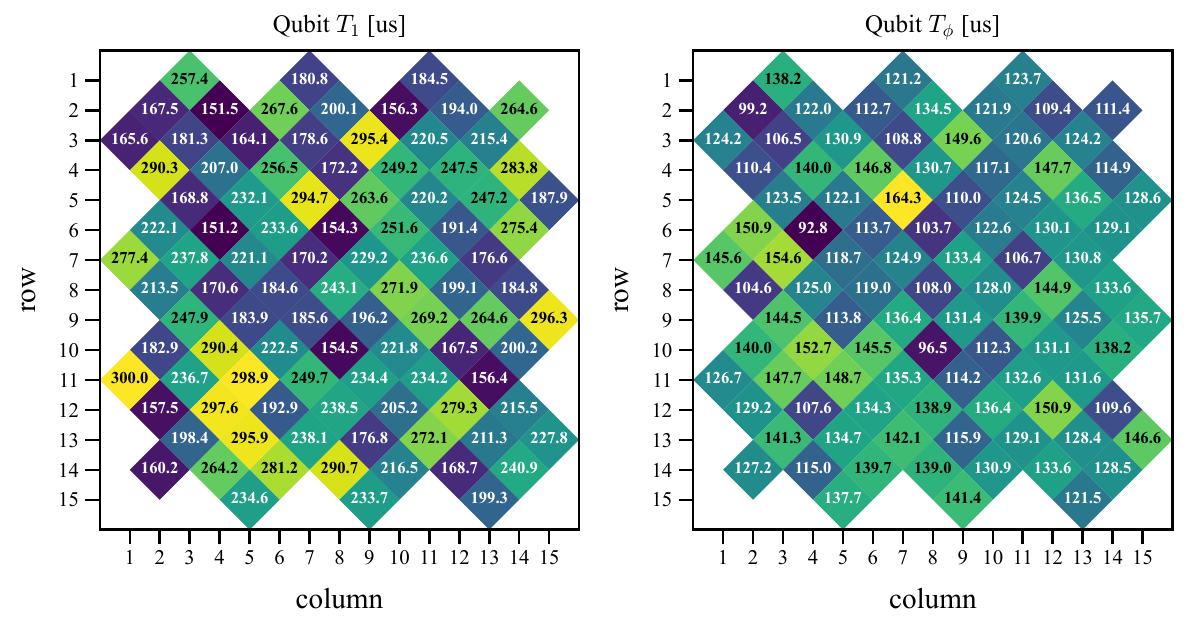}
    \caption{Heatmaps of single-qubit parameters. Left: qubit relaxation time ($T_1$). Right: qubit pure dephasing time ($T_\phi$).}
    \label{fig:1Q_params}
\end{figure}

The hardware parameters used in the simulations are chosen from representative characterization data on \texttt{ibm\_miami}. \cref{fig:graph_spec} summarizes the coherent parameters entering the Hamiltonian model. The residual $ZZ$ couplings $J_{ij}$ are extracted from diagnostic dynamical decoupling sequences: qubits or couplers are probed in an order determined by the device connectivity, so that non-overlapping neighborhoods can be characterized in parallel~\cite{brown2025efficient}. The measured $J_{ij}/2\pi$ values are mostly concentrated at the few-kHz to $\sim 10$ kHz scale, with a small number of larger outliers extending above $100$ kHz. For the distance-$7$ instance, we sample from the representative range listed in \cref{tab:noise_model} (excluding outliers). The drive-qubit detunings $\Delta_g/2\pi$ are centered near zero, with most values lying within the range used in the detuning sweep.

\cref{fig:1Q_params} shows the spatial variation of the single-qubit relaxation and pure dephasing times. These data set experimentally realistic scales for the simulated distance-$7$ surface-code instance. Together, \cref{fig:graph_spec,fig:1Q_params} motivate the parameter ranges listed in \cref{tab:noise_model}.

\section{Details on Pauli-twirling}
\label{app:Pauli-twirling}

The Gottesman-Knill theorem~\cite{gottesman1998heisenberg} states that any quantum circuit composed of Clifford unitaries, Pauli-state preparation, and Pauli-basis measurements can be simulated classically in polynomial time. This efficient stabilizer description is no longer directly applicable to general open-system dynamics, such as coherent always-on couplings, nonunital relaxation, general dissipative channels, and imperfect control Hamiltonians. A common way to recover a stabilizer-compatible description is to replace each finite-time noise process assigned to a circuit location by its Pauli-twirled approximation. This location-wise construction is an approximation when coherent always-on terms overlap multiple gates or idle periods, which is one reason we compare it against the continuous-time master-equation simulation.

Consider a completely positive and trace-preserving (CPTP) channel
\beq
    \rho \mapsto \mathcal{E}(\rho)
    =
    \sum_m E_m \rho E_m^\dagger ,
\eeq
where the $E_m$ are Kraus operators. Let
$\mathcal{P}_n=\{I,X,Y,Z\}^{\otimes n}$ denote the $n$-qubit Pauli operators without overall phases, so that $|\mathcal{P}_n|=4^n$. The Pauli twirl of $\mathcal{E}$ is defined as
\beq
    \tilde{\mathcal{E}}(\rho)
    =
    \frac{1}{|\mathcal{P}_n|}
    \sum_{A\in\mathcal{P}_n}
    A^\dagger \mathcal{E}(A\rho A^\dagger) A .
\eeq
Since the Pauli operators form an orthogonal operator basis under the Hilbert-Schmidt inner product, each Kraus operator can be expanded as
\beq
    E_m = \sum_{B\in\mathcal{P}_n} e_{mB} B ,
\eeq
with complex coefficients $e_{mB}$. Substituting this expansion into the twirled channel gives
\begin{align}
    \tilde{\mathcal{E}}(\rho)
    &=
    \frac{1}{|\mathcal{P}_n|}
    \sum_{m}
    \sum_{A,B,C\in\mathcal{P}_n}
    e_{mB} e_{mC}^*
    A^\dagger B A \rho A^\dagger C^\dagger A .
\end{align}
For any Pauli operators $A$ and $B$,
\beq
    A^\dagger B A = \chi_B(A) B ,
\eeq
where $\chi_B(A)= 1$ ($-1$) if $A$ commutes (anticommutes) with $B$. Therefore,
\beq\bal
    \tilde{\mathcal{E}}(\rho)
    &=
    \sum_{m}
    \sum_{B,C\in\mathcal{P}_n}
    e_{mB} e_{mC}^*\times\\
    &\quad \left[
    \frac{1}{|\mathcal{P}_n|}
    \sum_{A\in\mathcal{P}_n}
    \chi_B(A)\chi_C(A)
    \right]
    B\rho C^\dagger .
\eal\eeq
By character orthogonality over the Pauli group (modulo phases),
\beq
    \frac{1}{|\mathcal{P}_n|}
    \sum_{A\in\mathcal{P}_n}
    \chi_B(A)\chi_C(A)
    =
    \delta_{BC},
\eeq
all off-diagonal terms in the Pauli expansion vanish, leaving
\beq
    \tilde{\mathcal{E}}(\rho)
    =
    \sum_{m}
    \sum_{B\in\mathcal{P}_n}
    |e_{mB}|^2 B\rho B^\dagger= \sum_{B\in\mathcal{P}_n} p_B B\rho B^\dagger,
\eeq
with $p_B = \sum_m |e_{mB}|^2$. Since the original channel is trace preserving, the probabilities satisfy
\beq
    p_B \geq 0,
    \qquad
    \sum_{B\in\mathcal{P}_n} p_B = 1 .
\eeq
Thus, Pauli twirling replaces the original channel by a stochastic Pauli channel with probabilities given by diagonal Pauli-expansion weights $p_B$. In doing so, it removes 
off-diagonal Pauli coherences, coherent phase information, and affine drift terms for nonunital channels represented in Pauli transfer matrix form. These discarded features can affect syndrome statistics when errors propagate through a circuit, especially when coherent and dissipative mechanisms act simultaneously.

In the Clifford simulations used in the main text, the master-equation noise sources in \cref{tab:noise_model} are approximated by the  Pauli channels applied to the corresponding gate and idle intervals in the Stim circuit. This construction twirls the individual noise mechanisms assigned to each location; it is not the Pauli twirl of the full simultaneous finite-time master-equation evolution. Relaxation together with pure dephasing over a time interval $\Delta t$ is represented by the Pauli twirl of the combined $T_1/T_2$ channel,
\beq
    \mathcal{N}_{T_1,T_2}(\rho)
    =
    p_I\rho+p_X X\rho X+p_Y Y\rho Y+p_Z Z\rho Z .
\eeq
For this channel, $T_2$ denotes the effective transverse relaxation time defined, as in \cref{eq:T2}, by
$\frac{1}{T_2}=\frac{1}{2T_1}+\frac{1}{T_\phi}$.
Writing
\beq
    \eta_1=e^{-\Delta t/T_1},
    \quad
    \eta_2=e^{-\Delta t/T_2},
\eeq
the Pauli probabilities are
\beq\bal
    p_I &= \frac{1+2\eta_2+\eta_1}{4}, \\
    p_X &= p_Y = \frac{1-\eta_1}{4}, \\
    p_Z &= \frac{1-2\eta_2+\eta_1}{4}
         =
         \frac{1-e^{-\Delta t/T_2}}{2}-p_X .
\eal\eeq
This Pauli channel reproduces the diagonal Pauli-transfer contractions of the combined relaxation/dephasing process, but discards the nonunital affine drift associated with amplitude damping. The sampled parameters satisfy $T_2\leq 2T_1$, so $p_Z\geq 0$.
Residual $ZZ$ crosstalk generated by
\beq
    H_{ZZ}^{\rm xtalk}=J_{ij}Z_iZ_j
\eeq
over a time interval $\Delta t$ is twirled to
\begin{align}
    \mathcal{N}_{ZZ}^{\rm xtalk}(\rho)
    &= (1-p_{ZZ}^{\rm xtalk})\rho+p_{ZZ}^{\rm xtalk}P_{ij}\rho P_{ij}, \\
    p_{ZZ}^{\rm xtalk} &= \sin^2(J_{ij}\Delta t).
\end{align}
Here $J_{ij}$ is the angular-frequency coupling coefficient; when the experimentally quoted value is $J_{ij}/2\pi$, the factor of $2\pi$ is included before evaluating $J_{ij}\Delta t$. The Pauli operator $P_{ij}$ depends on the representation. In a sequential Stim circuit, the stochastic fault is inserted at the corresponding physical circuit location as the physical $Z_iZ_j$, and Stim propagates it through the subsequent ideal Clifford gates. Equivalently, in the toggling-frame the same fault becomes
\beq
    P_{ij}(t)
    =
    U_{\rm ideal}^{\dagger}(t)
    Z_iZ_j
    U_{\rm ideal}(t).
\eeq
For intervals with no preceding basis-changing ideal gates on either qubit, this reduces to $P_{ij}=Z_iZ_j$.

Single-qubit pulse-amplitude under-rotation for an implemented physical $R_X(\theta_H)$ is twirled to
\begin{align}
    \mathcal{N}_{1q,g}(\rho)
    &=
    (1-p_{1q})\rho+p_{1q}P_{1q,g}\rho P_{1q,g}, \\
    p_{1q}
    &=
    \sin^2 \left(\frac{\delta_g\theta_H}{2}\right).
\end{align}
Here $\delta_g$ is the fractional pulse-amplitude error, so the implemented angle is $(1-\delta_g)\theta_H$. For a sequential Stim circuit, the Pauli fault is inserted at the chosen physical circuit location with the Pauli index appropriate to that location. 
If the fault is inserted immediately after the physical $R_X$ pulse before subsequent ideal frame updates, then this index is $X$. If instead  the compiled Hadamard is represented as an ideal Clifford including trailing virtual-$Z$ updates before the noise insertion point, then the equivalent Pauli index is conjugated by those ideal frame updates. Virtual-$Z$ updates do not introduce independent stochastic faults, but they determine the Pauli frame in which pulse-amplitude errors are labeled.

For two-qubit pulse-amplitude under-rotation of an implemented $R_{ZZ}(\theta_{ZZ})$ gate, with fractional amplitude error $\delta_g$ so that the implemented angle is $(1-\delta_g)\theta_{ZZ}$, the residual unitary is a small $ZZ$ rotation. Its Pauli twirl is
\beq\bal
    \mathcal{N}_{ZZ}^{\rm gate}(\rho)
    &=
    (1-p_{ZZ}^{\rm gate})\rho
    +
    p_{ZZ}^{\rm gate} Z_iZ_j\rho Z_iZ_j, \\
    p_{ZZ}^{\rm gate}
    &=
    \sin^2 \left(\frac{\delta_g\theta_{ZZ}}{2}\right).
\eal\eeq
For a CZ-equivalent $R_{ZZ}(\pi/2)$ entangler in the convention $R_{ZZ}(\theta)=\exp(-i\theta Z_iZ_j/2)$, this gives
\beq
p_{ZZ}^{\rm gate}=\sin^2(\delta_g\pi/4) .
\eeq

When residual $ZZ$ crosstalk and $R_{ZZ}$ pulse under-rotation act during the same time interval, the Stim baseline applies their Pauli-twirled channels as separate stochastic faults. Under this convention, each noise mechanism is twirled locally before the resulting Pauli channels are composed. By contrast, a Pauli twirl of the combined coherent $ZZ$ error over the same interval would first add the coherent rotation angles and only then evaluate the corresponding $\sin^2$ Pauli probability.

No separate location-independent Pauli-twirled channel is assigned to drive-qubit detuning in the Stim reference model. A finite-time detuned pulse could be Pauli-twirled after computing its complete pulse-level error channel; 
here detuning is instead retained in the master-equation pulse Hamiltonian.

In the Stim implementation, these Pauli channels are inserted at the same gate and idle locations used in the circuit schedule. When multiple Pauli noise mechanisms are assigned to the same location, their individually twirled Pauli channels are composed to obtain the stochastic channel used at that location. Since Pauli channels commute as channels, the order of composition does not affect the resulting channel.

\section{Proof of the toggling-frame formula for $p_k^{\neq}$}
\label{app:p_k-neq}

We prove the identity claimed in \cref{eq:pneq_toggling_direct}. In this proof, $CZ$ denotes the ideal Clifford entangler represented in the circuit model; in the pulse model this Clifford is implemented by a CZ-equivalent $R_{ZZ}(\pi/2)$ pulse together with local virtual-$Z$ frame updates.

As in the main text, let
$A_k := Z_{a(k)}$
denote the final computational-basis readout Pauli on the measurement ancilla for check $k$, and let $S_k$ denote the corresponding data stabilizer. We use the syndrome-label convention in which a $+1$ readout of the measurement ancilla corresponds to a $+1$ eigenvalue of $S_k$.

To prove \cref{eq:pneq_toggling_direct}, we first need the following ideal-circuit identity:
\beq
 U_{\rm ideal}^{\dagger}(t_f) A_k U_{\rm ideal}(t_f) = A_k S_k .
 \label{eq:app_readout_identity}
\eeq
This is essentially the Heisenberg-picture Pauli-propagation form of the standard stabilizer-generator measurement circuit; see, e.g., Ref.~\cite[Ch.~10]{nielsen2010quantum}.
To prove \cref{eq:app_readout_identity}, let $\mathcal N(k)$ be the set of data qubits in check $k$. Choose a local data-basis Clifford $B_k$ such that
\beq
 S_k
 =
 B_k^{\dagger} \left(\prod_{q\in\mathcal N(k)} Z_q \right) B_k .
 \label{eq:app_stabilizer_basis_change}
\eeq
For a $Z$-type check, $B_k=I$; for an $X$-type check, $B_k=\prod_{q\in\mathcal N(k)}H_q$, where $H_q$ denotes a Hadamard applied to qubit $q$. Let
\beq
 E_k
 =
 \prod_{q\in\mathcal N(k)} CZ_{a(k),q}
\eeq
be the ideal entangling block coupling the ancilla to the data qubits in the check, where $CZ_{a(k),q}$ denotes the controlled-$Z$ between ancilla $a(k)$ and data qubit $q$. The corresponding ideal check-measurement block can be written as
\beq
 U_k^{\rm meas}
 =
 H_{a(k)} B_k^{\dagger} E_k B_k H_{a(k)} ,
 \label{eq:app_check_block}
\eeq
where the rightmost $H_{a(k)}$ prepares the ancilla in the $X$ basis before the entangling block, and the leftmost $H_{a(k)}$ maps the final ancilla $X$ information to computational-basis readout.

Propagating the final readout Pauli $A_k=Z_{a(k)}$ backward through this ideal block gives
\beq
\begin{aligned}
& \left(U_k^{\rm meas}\right)^{\dagger} A_k U_k^{\rm meas}\\
\quad &=
 H_{a(k)}B_k^{\dagger}E_k B_k H_{a(k)}Z_{a(k)} H_{a(k)} B_k^{\dagger} E_k B_k H_{a(k)}\\
\quad &=
 H_{a(k)} B_k^{\dagger} E_k X_{a(k)} E_k B_k H_{a(k)}\\
\quad &=
 H_{a(k)} B_k^{\dagger} \left(X_{a(k)} \prod_{q\in\mathcal N(k)} Z_q \right) B_k H_{a(k)}\\
\quad &=
 H_{a(k)} X_{a(k)} H_{a(k)} B_k^{\dagger} \left(\prod_{q\in\mathcal N(k)} Z_q \right) B_k\\
\quad &=
 Z_{a(k)}S_k= A_kS_k ,
\end{aligned}
\eeq
where the third equality uses the $X$-gate-propagation rule for a CZ gate [$CZ_{aq}X_a CZ_{aq} = X_aZ_q$], and the fourth equality just separates the ancilla part from the data part.

For the full ideal syndrome-extraction circuit, the same identity holds for the readout convention used here:
\beq
 U_{\rm ideal}^{\dagger}(t_f) A_k U_{\rm ideal}(t_f) = A_k S_k .
 \label{eq:app_readout_identity2}
\eeq
This is the Pauli-propagation statement that the final computational-basis readout of ancilla $a(k)$ reports the stabilizer $S_k$, up to the initial ancilla eigenvalue fixed by the preparation convention. In the full scheduled circuit, this identity can be verified by propagating $A_k$ backward through the complete Clifford schedule. Gates associated with other checks either do not act on ancilla $a(k)$ or implement quantum non-demolition (QND) measurements of stabilizers that commute with $S_k$, and hence do not change the final data stabilizer associated with this readout.

The ideal syndrome-extraction circuit is QND with respect to the stabilizers being measured, so it also preserves the measured data stabilizer:
\beq
 U_{\rm ideal}^{\dagger}(t_f) S_k U_{\rm ideal}(t_f) = S_k .
 \label{eq:app_stabilizer_invariance}
\eeq
For a single check block, this follows in the local basis of \cref{eq:app_stabilizer_basis_change} from commutation of the measured stabilizers with the CZ entangling block, together with cancellation of the ideal basis-change gates at the end of the block. For the full round, the same statement follows because all measured surface-code stabilizers commute and the scheduled ideal circuit implements their simultaneous QND measurement.

Combining \cref{eq:app_readout_identity2,eq:app_stabilizer_invariance}, we obtain
\beq
\begin{aligned}
 U_{\rm ideal}^{\dagger}(t_f) A_k S_k U_{\rm ideal}(t_f)
 &=
 \left[U_{\rm ideal}^{\dagger}(t_f) A_k U_{\rm ideal}(t_f) \right] \times\\
 &\quad \quad\left[U_{\rm ideal}^{\dagger}(t_f) S_k U_{\rm ideal}(t_f) \right]\\
 &= (A_kS_k)S_k=A_k .
\end{aligned}
 \label{eq:app_AS_simplification}
\eeq

Now consider the disagreement probability. In the rotating control frame, the event that the ancilla readout disagrees with the data stabilizer is represented by the projector
\beq
 \Pi_k^{\neq} = \frac{I-A_kS_k}{2},
\eeq
because $A_kS_k=+1$ when the two signs agree and $A_kS_k=-1$ when they disagree. Thus, in the rotating control frame before the final toggling-frame change of variables,
\beq
 p_k^{\neq} = \Tr\left[\Pi_k^{\neq}\rho(t_f)\right].
\eeq
Using the toggling-frame relation
\beq
 \tilde\rho(t_f) = U_{\rm ideal}^{\dagger}(t_f) \rho(t_f) U_{\rm ideal}(t_f),
\eeq
cyclicity of the trace gives
\beq
\begin{aligned}
 p_k^{\neq}
 &=
 \Tr\left[
  U_{\rm ideal}^{\dagger}(t_f) \Pi_k^{\neq} U_{\rm ideal}(t_f) \tilde\rho(t_f) \right]\\
 &=
 \Tr\left[\frac{I-U_{\rm ideal}^{\dagger}(t_f)A_k S_k U_{\rm ideal}(t_f)}{2}\tilde\rho(t_f)
 \right].
\end{aligned}
\eeq
Using \cref{eq:app_AS_simplification}, this proves \cref{eq:pneq_toggling_direct}.

\bibliography{references}
\bibliographystyle{apsrev4-2}

\end{document}